\documentclass[usenatbib]{mn2e}
\usepackage{graphicx}
\usepackage{graphics}
\usepackage{epsf}
\usepackage{epstopdf}
\usepackage{breqn}
\usepackage{amssymb}
\newcommand{\msol}{{\rm M_{\odot}}}
\newcommand{\apjs}{ApJS}
\newcommand{\apj}{ApJ}
\newcommand{\apjl}{ApJ}
\newcommand{\mnras}{MNRAS}

\newcommand{\aap}{A\&A}
\newcommand{\araa}{ARA\&A}
\newcommand{\physrep}{PhR}
\newcommand{\nat}{Nature}
\newcommand{\abs}[1]{\lvert#1\rvert}
\title[2-D MHD Simulations of Galaxy Clusters]{2-D Magnetohydrodynamic Simulations of Induced Plasma Dynamics in the Near-Core Region of a Galaxy Cluster}
\author[I. G. Mikellides, K. Tassis and H. W. Yorke]{I. G. Mikellides,$^{1}$\thanks{E-mail:
Ioannis.G.Mikellides@jpl.nasa.gov} K. Tassis,$^{1}$ and H. W. Yorke,$^{1}$\\
$^{1}$Jet Propulsion Laboratory, California Institute of Technology, Pasadena, CA, 91109\\}

\begin{document}

\date{Accepted 2010 Month ??. Received 2010 Month ??; in original form 2010 Month ??}

\pagerange{\pageref{firstpage}--\pageref{lastpage}} \pubyear{2002}

\maketitle

\label{firstpage}

\begin{abstract}
The mechanisms that maintain thermal balance in the intracluster medium (ICM) and produce the observed spatial distribution of the plasma density and temperature in galaxy clusters remain a subject of debate.  We present results from numerical simulations of the cooling-core cluster A2199 produced by the two-dimensional (2-D) resistive magnetohydrodynamics (MHD) code MACH2. In our simulations we explore the effect of anisotropic thermal conduction on the energy balance of the system. The results from idealized cases in 2-D axisymmetric geometry underscore the importance of the initial plasma density in ICM simulations, especially the near-core values since the radiation cooling rate is proportional to ${n_e}^2$. Heat conduction is found to be non-effective in preventing catastrophic cooling in this cluster.  In addition we performed 2-D planar MHD simulations starting from initial conditions deliberately violating both thermal balance and hydrostatic equilibrium in the ICM,  to assess contributions of the convective terms in the energy balance of the system against anisotropic thermal conduction. We find that in this case work done by the pressure on the plasma can dominate the early evolution of the internal energy over anisotropic thermal conduction in the presence of subsonic flows, thereby reducing the impact of the magnetic field. Deviations from hydrostatic equilibrium near the cluster core may be associated with transient activity of a central active galactic nucleus and/or remnant dynamical activity in the ICM and warrant further study in three dimensions. 
\end{abstract}


\section{Introduction}\label{sec:Intro}
Clusters of galaxies are the largest structures in the universe. The majority of their barionic mass is in the form of a hot plasma filling the space between galaxies, the intracluster medium (ICM). The cooling time of the ICM near the center of the clusters due to radiative losses via X-ray emission is shorter than the Hubble time for most clusters (\citealt{1986RvMP...58....1S}; \citealt{1992MNRAS.258..177E}; Peres et al. 1998; Sanderson et al. 2006). In the absence of heating the short cooling times (as short as $\sim$0.1 Gyr) are expected to lead to cooling of the ICM plasma and its inflow towards the center of the cluster at a rate that was estimated to be as large as $\sim10^3$ $\msol$ $yr^{-1}$ (Cowie \& Binney 1977; Fabian \& Nulsen 1977; Mathews \& Bregman 1978; for a review, see Fabian 1994). However, high resolution X-ray spectroscopy from \textit{XMM-Newton} and \textit{Chandra} have failed to find evidence of such ``cooling flows'' (for a review see Peterson \& Fabian 2006), strongly suggesting that one or more heating mechanisms must be balancing radiative cooling in galaxy clusters.

A wide variety of heating mechanisms have been investigated, including i) heat transport from the outer regions of the cluster to the central cooling gas by conduction (e.g. Binney \& Cowie 1981; Tucker \& Rosner 1983; Bertschinger \& Meiksin 1986; Bregman \& David 1988; Gaetz 1989; Rosner \& Tucker 1989; David et al. 1992; Pistinner \& Shaviv 1996; Narayan \& Medvedev 2001; Voigt et al. 2002; Fabian et al. 2002; Zakamska \& Narayan 2003; Kim \& Narayan 2003a; Markevitch et al. 2003; Dolag et al. 2004; Xiang et al. 2007; Pope et al. 2006; Balbus \& Reynolds 2008; Bogdanovi{\'c} et al. 2009; Parrish et al. 2009; Parrish et al. 2010); ii) energy  injection from active galactic nuclei (AGN) via jets, bubbles, radiation and sound waves (e.g. Binney \& Tabor 1995; Ciotti \& Ostriker 2001; Br{\"u}ggen \& Kaiser 2002; Omma et al. 2004; Ruszkowski et al. 2004a,b; Voit \& Donahue 2005; Fabian et al. 2005; Nusser et al. 2006; Mathews et al. 2006; Chandran \& Rasera 2007; McCarthy et al. 2008;  Br{\"u}ggen \& Scannapieco 2009); iii) heating due to galaxy motions via dynamical friction (e.g. Schipper 1974; Miller 1986; Just et al. 1990; Fabian 2003; El-Zant et al. 2004; Kim et al. 2005; Dekel \& Birnboim 2008); iv) turbulent mixing (e.g. Loewenstein \& Fabian 1990; Deiss \& Just 1996; Ricker \& Sarazin 2001; Kim \& Narayan 2003b; Cho et al. 2003; Dennis \& Chandran 2005; ZuHone et al. 2009); v) magnetic field reconnection (Soker \& Sarazin 1990); vi) viscous heating regulated by pressure anisotropy (e.g. \citealt{2010arXiv1003.2719K}); or vii) a combination of these (e.g. Ruszkowski \& Begelman 2002; Brighenti \& Mathews 2002; Brighenti \& Mathews 2003; Voigt \& Fabian 2004; Fujita \& Suzuki 2005; Pope et al. 2006; Conroy \& Ostriker 2008; Guo et al. 2008).

Due to the extremely high Hall parameter of ICM plasmas the impact of the magnetic field on the thermal balance of the ICM has been of particular interest; extensive multi-dimensional magnetohydrodynamic (MHD) simulations have been employed in the last several years to assess anisotropic effects. \citet{2008ApJ...677L...9P} performed two-dimensional (2-D) and 3-D simulations with the MHD code \textit{Athena} (\citealt{2008JCoPh.227.4123G, 2008ApJS..178..137S}) focusing on the evolution and saturation of the heat-flux-driven buoyancy instability (HBI) (\citealt{2008ApJ...673..758Q}) for various magnetic field strengths. \citet{2009ApJ...704..211B} also performed 3-D simulations with \textit{Athena}. They carried out pre-simulations, whereby starting with an isothermal ICM they allowed radiative cooling without thermal conduction to obtain a cooling core state, and then used \textit{it} as the ``initial condition'' for their self-consistent MHD simulations. For the initial magnetic field the authors' numerical experiments included a case of rotational field lines. They concluded that thermal conduction alone cannot be the solution to the cooling flow problem. Moreover, they hypothesized that the nature of the AGN feedback is one of ``stirring'' rather than heating, that is, AGN activity periodically disrupting the azimuthal field structure allowing thermal conduction to sporadically heat the core. The level of effectiveness of ``stirring'' however remains an open question. For example, cooling-core clusters like the A2199 contain prominent metal gradients that could prevent violent processes from stirring the ICM. In a similar study \citet{2009ApJ...703...96P} performed a wide range of numerical studies of the A2199 cluster with tangled and radial initial magnetic field topologies and strengths of 1 nG to 1$\mu$G, extending the parameter study of \citet{2009ApJ...704..211B}. They demonstrated that the effective reduction of the anisotropic heat flux was less than 10\% of the Spitzer value due to a re-orientation of the magnetic field lines relative to the radial direction induced by the HBI. This reduction was found to occur withing several times the HBI growth time ($\sim$0.125 Gyr for A2199), leading to a cooling catastrophe in less than $\sim$3 Gyr. The authors found similar trends at different initial magnetic field strengths and topologies. Starting with lower initial core densities and isothermal initial temperature the cooling catastrophe was significantly delayed to $\sim$7 Gyr. The cooling catastrophe was also delayed when a central heating source was assumed via an idealized model of the heating luminosity within a radius of 20 kpc. The authors concluded (at least for A2199) that in the absence of AGN feedback and/or strong magnetic fields near the galaxy cluster core, thermal conduction alone cannot be responsible for balancing the radiative losses in the ICM, in agreement with \citet{2009ApJ...704..211B}. Similar results about the reduction and evolution of the effective heat flux were obtained by \citet{2010ApJ...713.1332R} using the 3-D MHD code \textit{FLASH}.\footnote{http://flash.uchicago.edu}

There is a large body of numerical work on various aspects of AGN-ICM interactions that has spanned purely hydrodynamic models (e.g. \citealt{2001ApJ...554..261C, 2001MNRAS.325..676B, 2002Natur.418..301B, 2002MNRAS.332..271R, 2004MNRAS.348.1105O, 2005A&A...429..399Z, 2006ApJ...645...83V}), as well as MHD models (e.g. \citealt{2004ApJ...615..675R, 2004ApJ...601..621R, 2005ApJ...624..586J, 2010ApJ...710..743D}). Yet, a conclusive picture regarding the cooling flow problem based on numerical simulations appears to remain elusive. For example, \citet{2006ApJ...645...83V} simulated direct injection of a supersonic jet into the ICM atmosphere. However, this model failed to maintain long-term thermal balance in the cluster, leading to hypotheses of heating by the dissipation of sound waves or the decay of global ICM modes (e.g \citealt{2004MNRAS.348.1105O}). Also, \citet{2010ApJ...710..743D} argue based on comparisons with experimental data that AGN outflows are strongly decelerated, generating turbulent sonic or subsonic flows due to their interaction with the surrounding medium, which support such hypotheses.

In this paper we present results from a series of 2-D numerical experiments aimed at assessing the driving physics and related sensitivities in a galaxy cluster. We have chosen to simulate A2199, a cooling-core cluster that has been simulated by other authors (e.g. \citealt{2003ApJ...582..162Z, 2009ApJ...703...96P, 2010ApJ...713.1332R}), and for which observed data for the electron number density and temperature exist at a relatively close distance ($R \sim$1 kpc) from the cluster center. The density at the core is an important parameter in MHD parameter studies that aim to assess mechanisms associated with the sustainment (or collapse) of the thermal balance in a cluster, since the radiative cooling rate depends strongly on it ($\propto n_e^2$). Because A2199 is a cluster that is often modeled by MHD simulations, we perform here a series of idealized numerical experiments in 2-D axisymmetric geometry to assess the sensitivity of the collapse rate on the near-core density, using two different spatial profiles that bounded the observational data (\citealt{2002MNRAS.336..299J}).

In 3-D MHD simulations it is common to prescribe initial profiles that are in hydrostatic equilibrium. Often, the observational data guide solutions to the hydrostatic equation and then these solutions are prescribed as the ``initial conditions'' of the ICM. MHD simulations then proceed to seek of mechanisms that maintain thermal balance in the cluster. In this study we perform a series of numerical experiments that do not begin with solutions in hydrostatic equilibrium, to assess dynamical effects in the energy balance of the system. Our simulations include a range of initial magnetic field profiles (similar to those used by \citealt{2009ApJ...703...96P} and \citealt{2009ApJ...704..211B}) and quantify the self-consistent evolution of the ICM in 2-D planar geometry.

The paper is organized as follows. Section \ref{sec:Method} describes the computational methodology and governing laws used to perform the numerical simulations. Initial and boundary conditions for the cluster are also given in this section. Section \ref{sec:NumExperiments} presents our simulation results in 2-D axisymmetric (r,z) and in 2-D planar (x,y) physical domains. We summarize the results and provide our conclusions in Sections \ref{sec:Discussion} and \ref{sec:Conclusions}, respectively.

\section{method}\label{sec:Method}
The numerical simulations have been performed with the \textbf{M}ultiblock \textbf{A}rbitrary \textbf{C}oordinate \textbf{H}ydromagnetic (MACH) code (\citealt{1998JCoPh.140..148P}). MACH$x$ (where $x=2$ or 3) are time-dependent, resistive, compressible MHD solvers of the Arbitrary Lagrangian Eulerian (ALE) variety that were developed by the Center for Plasma Theory and Computation at the Air Force Research Laboratory, and NumerEx Inc., to support applications with non-ideal MHD physics in complex geometries. MACH2 and MACH3 are the two-and-one-half-dimensional (2$\frac{1}{2}$-D) and 3-D versions of the solvers, respectively. Over the last three decades they have been upgraded significantly through the contributions of many scientists in the United States. Although the majority of the work with MACH today is in the area of pulsed power, their application base has expanded to multiple areas in engineering. However, to the best of our knowledge, they have not yet been used to study problems in astrophysics. In order to introduce these codes to the astrophysical community, we provide a more detailed description of the codes and their applications in Appendix A.

MACH2 is the version used in the present study. A feature of MACH2 that has been particularly useful in our studies is its modularity with the terms in the MHD equations, which allowed us to perform instructive comparisons between the various simulation physics in the evolution of the ICM. The numerical experiments have been carried out using a reduced set of the full MHD conservation laws of MACH (see Appendix A). Simulations in only two dimensions allow a wide range of highly-resolved cases at reduced computational time, but at the expense of excluding processes in the evolution of the ICM that are inherently 3-D. Nevertheless, the range of physics cases examined has provided us with a useful set of results, especially in regards to the ``initial state'' of the ICM and the potential importance of low-speed hydrodynamics near the center of the cluster. These 2-D results now also form the basis of our future simulations with MACH3. 

\subsection{Governing equations and numerical approach}\label{subsec:Equations}
The ICM simulations have been performed using the following reduced set of the full MACH conservation laws, representing ideal MHD physics in the presence of anisotropic thermal conduction and radiative cooling:
\begin{eqnarray}
%Fluid continuity
\frac{\partial \rho}{\partial t}=-\nabla\cdot\left(\rho \textbf{\textit{v}}\right)
\label{ICMcontinuity}
\end{eqnarray}
\begin{eqnarray}
%Fluid momentum
\rho\frac{\partial \textbf{\textit{v}}}{\partial t}=-\rho \textbf{\textit{v}}\cdot\nabla \textbf{\textit{v}}-\nabla p+{\mu_0}^{-1}\left(\nabla\times \textbf{\textit{B}}\right)\times \textbf{\textit{B}}+\rho \textbf{\textit{g}}
\label{ICMmomentum}
\end{eqnarray}
\begin{eqnarray}
%specific internal energy
\rho\frac{\partial \epsilon}{\partial t}=-\rho \textbf{\textit{v}}\cdot\nabla \epsilon-p\nabla \cdot \textbf{\textit{v}}-\nabla\cdot\textbf{\textit{q}}-\Phi_{eR}
\label{ICMenergy}
\end{eqnarray}
\begin{eqnarray}
%Magnetic induction
\frac{\partial \textbf{\textit{B}}}{\partial t}=\nabla\times\left(\textbf{\textit{v}}\times \textbf{\textit{B}}\right).
\label{ICMinduction}
\end{eqnarray}
In equations (\ref{ICMcontinuity})-(\ref{ICMinduction})
 $\rho$ is the mass density of the plasma fluid,
 $\textbf{\textit{v}}$ is the fluid velocity field,
 $\textbf{\textit{B}}$ is the magnetic induction field,
 $\epsilon$ is the internal specific energy,
 $\textbf{\textit{q}}$ is the conductive heat flux, 
 $p$ is the thermal pressure and
 $\textbf{\textit{g}}$ is the gravitational acceleration.

Analytic models or semi-empirical tabular equations of state (EOS) and transport coefficients may be used in MACH by accessing the \textit{sesame} library \footnote{http://t1web.lanl.gov/doc/SESAME\_3Ddatabase\_1992.html} (see Appenix A for more information about the library) during each computational cycle. For the cluster simulations we have used an ideal-gas EOS with a specific heat ratio $\gamma=$ 5/3,
%%%%%%%%%%%%%%%%%%%%%%%%
\begin{equation}
  \epsilon= \frac{p}{\rho (\gamma-1)}.
\end{equation}
Also, an analytic model of the anisotropic thermal conductivity coefficients has been used, which incorporates the full Braginskii coefficients (\citealt{1965RPP}) for the parallel ($\kappa_{e//}$) and perpendicular ($\kappa_{e\bot}$) conductivities
\begin{equation}
 \bmath{\bar{\bar{\kappa}}}_e=\kappa_{e//}\bmath{\hat{e}}_B\bmath{\hat{e}}_B+\kappa_{e \bot}     \left({\bar{\bar{\textbf{I}}}}-\bmath{\hat{e}}_B\bmath{\hat{e}}_B\right)
\label{eqn:thcondtensor}
\end{equation}
where $\bmath{\bar{\bar{\kappa}}}_e$ is the thermal conductivity tensor, $\bmath{\hat{e}}_B$ is the unit vector in the direction of the magnetic field and $\bar{\bar{\textbf{I}}}$ is the delta tensor; $\bmath{\hat{e}}_B\bmath{\hat{e}}_B$ is the magnetic field unit dyad. The Braginskii coefficients for $\bmath{\bar{\bar{\kappa}}}_e$ are provided in Appendix A. The conduction heat flux is given by
\begin{equation}
  \textbf{\textit{q}}=-\bmath{\bar{\bar{\kappa}}}_e\cdot \nabla T_e.
  \label{eqn:heatflux}
\end{equation}

In the relevant range of temperatures ($\sim$1-10 keV) the plasma in the cluster cools primarily by thermal bremsstrahlung at a rate given by (in units of $J/m^3/s$) $\Phi_{Tb}=1.4\times 10^{-28}\bar{g}_b {n_e}^2 {T_e}^{1/2}$ (e.g. \citealt{1979rpa..book.....R}) with $\bar{g}_b$ denoting the velocity-averaged Gaunt factor and $n_e$ is the electron number density. MACH2 uses a mean Planck opacity $\chi_p$ (with dimensions of area cross-section/unit mass, so a photon mean-free-path is $1/\chi_p$). The opacity values are taken from the \textit{sesame} library. In the optically-thin limit radiation energy is removed from a MACH2 computational element at a volumetric cooling rate of
\begin{equation}
  {\Phi_{eR}\\=a c_{\ell}\rho\chi_p\\{T_e}^4},
\end{equation}
where \textit{a} is Stefan's constant and $c_{\ell}$ is the speed of light in vacuum. A comparison of the cooling rates $\Phi_{Tb}$ and $\Phi_{eR}$ is provided in Appendix A.

The gravitational acceleration is obtained from Poisson's equation
\begin{equation}
{\nabla\cdot\textbf{g}\approx -4 \pi G\rho_{DM}}
\label{PoissonsGrav}
\end{equation}
where $\rho_{DM}$ is the mass density of the dark matter. Newton's gravitational constant is denoted by \textit{G}. The present study accounts only for the dark matter density, which is prescribed based on a Navarro-Frenk-White (NFW) profile with a soft core (\citealt{1997ApJ...490..493N}). The density profile may be expressed in non-dimensional form as follows:
\begin{equation}
\bar{\rho}_{DM} \left(\bar{r}\right)=\frac{1}{\left(\bar{r}+\bar{r_c}\right) \left(\bar{r}+1\right)^2}
\label{rhoDM}
\end{equation}
where $\bar{\rho}_{DM}\equiv 2\pi {R_s}^3\rho_{DM}/M_0$, $\bar{r}\equiv R/{R_s}$ and $\bar{r}_c\equiv {R_c}/{R_s}$ with $R$ denoting the radius measured from the cluster origin ($r=z=0$  or $x=y=0)$. The normalization mass is taken to be $M_0=3.8\times 10^{14} M_\odot$ from \citet{2003ApJ...582..162Z} and it has been kept fixed throughout all simulations presented herein. The standard scale radius for A2199, $R_s=$390 kpc, is taken also from \citet{2003ApJ...582..162Z} who inferred its value from observations of the outer plasma temperature. The value of the core radius $R_c$ is uncertain. For the ten clusters studied by Zakamska $\&$ Narayn the authors' ``best-fit'' solutions correspond to values of $\bar{r}_c=$ 0 and 1/20. For A2199 they chose ${R_c}=0$, whereas \citet{2010ApJ...713.1332R} chose ${R_c}=$20 kpc. We found that our main conclusions were not affected by this value and present our results for ${R_c}=$20 kpc. The solution of equation (\ref{PoissonsGrav}) for the gravitational acceleration is spherically symmetric, $\textbf{g}=g(R)\bmath{\hat{e}}_R$, where $\bmath{\hat{e}}_R$ is the position unit vector defined relative to the cluster origin. In non-dimensional form the analytical solution to equation (\ref{PoissonsGrav}) is given by:
\begin{eqnarray}
\bar {g}\left(\bar{r}\right) = &-&\frac{1}{\bar{r}\left(\bar{r}+1 \right)\left(\bar{r}_c-1\right)}+\frac{2 \bar{r}_c-1}{\bar{r}^2 \left(\bar{r}_c-1  \right)^2}\ln\left({\bar{r}+1}\right)\nonumber \\
 &-&\frac{{\bar{r}_c}^2}{\bar{r}^2 \left(\bar{r}_c-1  \right)^2}\ln\left({\frac{\bar{r}}{\bar{r}_c}+1}\right), 
\label{NFWgravaccel}
\end{eqnarray}
where $\bar{g}\equiv {R_s}^2 g/2GM_0$. Based on estimated values of the NFW characteristic radii and normalized mass for A2199 we find $\rho_{DM}\approx 1.3\times 10^{-24} g/cm^3$ at $R=$0.95 kpc. At the same location the observed electron number density is 0.115 $cm^{-3}$ and is the highest value we have used in our initial conditions. This gives $\rho\approx 0.2\times 10^{-24} g/cm^3$ for the plasma assuming that the molecular weight per electron is $\mu_e=$1.18. The latter assumed that the hydrogen and helium fractions are $X=0.7$ and $Y=0.28$, respectively (\citealt{2003ApJ...582..162Z}). Thus, our assumption in equation (\ref{PoissonsGrav}) about the dominance of the dark matter density in the determination of the gravitational acceleration appears acceptable.

Equations (\ref{ICMcontinuity})-(\ref{ICMinduction}) are solved in a fractional time-split manner. The Lagrangian time advance of the mass density, velocity and magnetic vector fields are computed implicitly on a computational domain that, unless otherwise noted, is discretized by a uniform mesh that consists of square computational elements (or ``cells''). Simulations in both 2-D axisymmetric (r,z) and 2-D planar (x,y) coordinate systems have been performed but no mesh movement and/or adaptation has been invoked. Spatial discretizations are obtained using finite volume differencing. The radiation cooling algorithm advances the internal energy explicitly, whereas thermal conduction is performed implicitly. Because we have assumed that the gravitational acceleration is dependent only on the dark matter density, $\textbf{\textit{g}}$ is prescribed using equation (\ref{NFWgravaccel}) and is held fixed throughout each simulation.

\subsection{Initial and boundary conditions}\label{subsec:Conditions}
\subsubsection{Initial conditions for the plasma density and temperature}\label{subsec:IConditions}
\begin{figure}
%\begin{center}
%\includegraphics[trim = <trim from left> <from bottom> <from right> <from top>, clip, <other options>]{<image filename>}
%\includegraphics[trim = 5mm 0mm 7.5mm 4mm, clip, scale=0.38]{InitialProfiles.png}
\includegraphics[trim = 5mm 0mm 7.5mm 4mm, clip, scale=0.48]{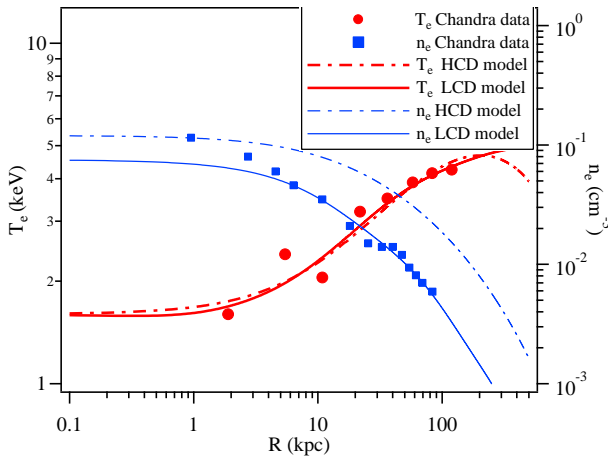}
 \caption{\textit{Chandra} data (\citealt{2002MNRAS.336..299J}, also in \citealt{2003ApJ...582..162Z}) and initial profiles used in the A2199 galaxy cluster simulations.}
\label{InitialProfiles}
%\end{center}
\end{figure}

Several combinations of initial conditions for the scalars and vectors have been used throughout this study, all associated with the A2199 galaxy cluster. Specifically, for the number density and temperature of the plasma we employ two different profiles to illustrate the sensitivity of the ICM simulations on the assumed initial conditions. The first profile is a fit of the Zakamska $\&$ Narayan ``best'' solution to an idealized set of four, time-independent, ordinary differential equations that account for thermal conduction and radiation cooling only, in hydrostatic equilibrium. The authors sought the solution that best fits the \textit{Chandra} observations (\citealt{2002MNRAS.336..299J}) by adjusting the thermal conduction heat flux factor  \textit{f}, defined here as
\begin{equation}
 \textit{f}\equiv\frac{\bmath{q}\cdot\bmath{\hat{e}}_R}{q_{sp}}
\end{equation}
where $q_{sp}\equiv \kappa_{e//}\nabla T_e\cdot\bmath{\hat{e}}_R$. The Spitzer-Braginskii thermal conductivity is ${\kappa_{e//}=\kappa_{sp}=3.16 {k_B}^2 n_e T_e\tau_e/m_e}$, where $\tau_e$ is the (classical) e-i collision time. The authors showed that for five out of the ten clusters studied a pure conduction model balancing radiative cooling, i.e. 
\begin{eqnarray}
\nabla\cdot \left(\textit{f} \kappa_{sp} \nabla T_e\right)-\Phi_{eR}=0,
\label{ICMenergyIdeal}
\end{eqnarray}
was plausible if the factor \textit{f} was reduced from unity by more than 50\%. Because this best-fit solution yields an electron number density near the core which is about 35\% lower than the observed value, we shall label it ``LCD'' (Low Core Density). The LCD profiles are plotted in Fig. \ref{InitialProfiles} against $\textit{Chandra}$ data (\citealt{2002MNRAS.336..299J}, also in \citealt{2003ApJ...582..162Z}). 

Some MHD simulations in the literature have employed initial density profiles that underestimate the observed electron number density by factors of 5-6 near the cluster core (e.g. see \citet{2009ApJ...703...96P} and \citet{2010ApJ...713.1332R}). To assess the importance of the core conditions on the evolution of the ICM we constructed a second set of initial conditions in a manner similar to that followed by \citet{2010ApJ...713.1332R}, who assumed hydrostatic equilibrium and a polytropic EOS with a prescribed profile for the entropy $S$ (e.g. \citealt{2009ApJS..182...12C}). We write the entropy as a function of radius in non-dimensional form:
\begin{equation}
 \bar{S}\left(\bar{r}\right)=\frac{\bar{T}}{\bar{\rho}^{\gamma-1}} =1+\beta \bar{r}/c,
\end{equation}
where the barred quantities are defined as $\bar{S} \equiv S/S_c$, $\bar{T} \equiv T_e/T_c$, $\bar{\rho} \equiv \rho/\rho_c$, $\beta \equiv (T_{200}/T_{c})(\rho_c/\rho_{200})^{\gamma-1}$ and $c=R_{200}/R_s$ is the concentration parameter. Here, variables carrying subscript ``c'' represent values at the cluster center and those with a subscript ``200'' denote values at a radius where the mean density of the dark matter is 200 times the critical density of the universe at the redshift of the cluster $\rho_{crit}(z)$. 

The hydrostatic equation is given by
\begin{equation}
 \nabla p=\rho \textbf{g}.
 \label{hydrostaticeqn}
\end{equation}
Assuming spherical symmetry, equation (\ref{hydrostaticeqn}) may be written in terms of the non-dimensional density and entropy as follows:
\begin{equation}
 \frac{d\bar{\rho}}{d\bar{r}}=\frac{1}{\gamma}\frac{\bar{\rho}}{\bar{S}}\left(\frac{C\bar{g}}{\bar{\rho}^{\gamma-1}}-\frac{d\bar{S}}{d\bar{r}}\right)
\label{hydrostaticeqn1D}
\end{equation}
and, given $\bar{S} (\bar{r})$, yields the density as a function of radius. The gravitational acceleration is taken from equation (\ref{NFWgravaccel}). The constant on the right of equation (\ref{hydrostaticeqn}) is $C \equiv 2GM_0\mu m_u/R_sk_BT_c$ ($\mu$ denotes the mean molecular weight, $m_u$ is the atomic mass unit and $k_B$ is Boltzmann's constant). Equation (\ref{hydrostaticeqn}) has been solved numerically to produce a second set of cluster profiles by assuming the following parameters: $T_c=$1.6 keV, $T_{200}=$4 keV, $\rho_c=23.5 \times 10^{-26} g/cm^{-3}$, $\rho_{200}=0.054 \times 10^{-26} g/cm^{-3}$, $c=$4.378, $\gamma=$5/3 and $\mu=$0.62. The profiles are plotted in Fig. \ref{InitialProfiles} and shall be denoted ``HCD'' (High Core Density) hereinafter. Although neither the LCD nor the HCD profiles represent accurately the density observations, they were implemented in this manner deliberately as limiting solutions that bounded the full range of the data, within the constraints of the hydrostatic equilibrium condition, while illustrating the sensitivity of the simulation results on the density.
\subsubsection{Initial conditions for the magnetic field}\label{subsec:IBConditions}
An extensive body of work exists on the magnetic field structure in galaxy clusters covering both cooling-core and non-cooling-core clusters. Faraday Rotation Measures towards extended radio sources in and behind galaxy clusters have demonstrated the turbulent nature of the magnetic field there (e.g. see \citealt {2010arXiv1007.5207G}, \citealt {2010A&A...514A..71V}, \citealt {2010A&A...514A..50G}, \citealt {2010A&A...513A..30B}, \citealt {2008A&A...483..699G}, \citealt {2006AN....327..533L}, \citealt {2005A&A...434...67V}, \citealt {2004A&A...424..429M}). In general it has been found that a power-law distribution of their fluctuation spectrum and some radially-declining profile of their average strength represent well these magnetic field structures. Nearly all hydrodynamical simulations within a cosmological context predict around 10\% of turbulence within the ICM (e.g. see \citealt {1998ApJ...495...80B}, \citealt {2003AstL...29..791I}, \citealt {2004MNRAS.351..505G}, \citealt {2006MNRAS.369L..14V}, \citealt {2008MNRAS.388.1079I}, \citealt {2010arXiv1001.1170P}), and are in agreemnt with observations (e.g. see \citealt {2004A&A...426..387S}, \citealt {2004MNRAS.347...29C}). Consequently, since the magnetic field is frozen within such turbulent ICM, cosmological MHD simulations predict that the magnetic field has a turbulent spectrum (specific examples can be found in \citealt {2002A&A...387..383D}, \citealt {2005AN....326..610B}) with no exception when studying a sample of clusters or during the time evolution of a single cluster. The reason for this is that clusters forming in a cosmological context contain accreting matter, part of which forms by small-scale-structure mergers. Therefore, clusters may be populated by thousands of sub-structures in the form of galaxies or former groups of galaxies that move through the cluster with speeds that are typcially on the order of a thousand km/s, permanently perturbing the underlying potential. It has been shown that such gravitational perturbations alone can produce significant motions within the core of clusters (e.g. see \citealt {2006ApJ...650..102A}, \citealt {2010IAUS..262..420R}, \citealt {2010ApJ...717..908Z}). Some very small fraction of these sub-structures contains also gaseous atmospheres or disks capable of steering the ICM directly.

To assess the impact of initial magnetic field topologies on the evolution of the effective thermal conduction in the ICM we employed here the following three initial magnetic field configurations: (1) a turbulent field (using random number deviates), (2) a purely rotational field and (3) a purely radial field. The intent of the last two highly-idealized configurations was to assess the sensitivity of the anisotropic heat flux reductions, and have served as the two limiting cases of our parameter studies on magnetic fields. In all three cases the out-of-plane component of the magnetic field, $B_{\theta}$ in the 2-D axisymmetric cases and $B_z$ in the planar cases, was set to zero.

\begin{figure*}
\includegraphics[trim = 1mm 1mm 1mm 1mm, clip, scale=0.44]{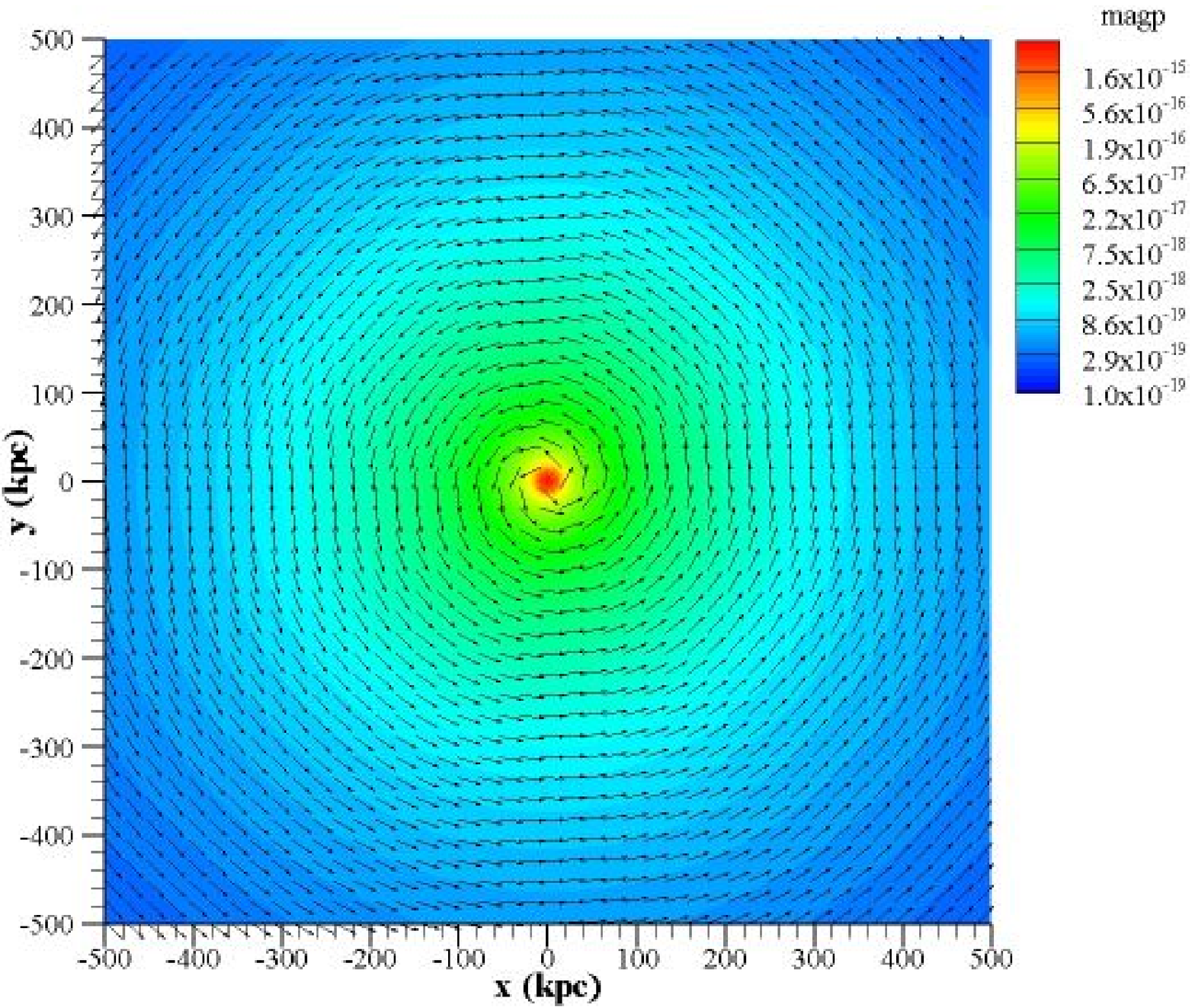}
\includegraphics[trim = 1mm 1mm 1mm 1mm, clip, scale=0.44]{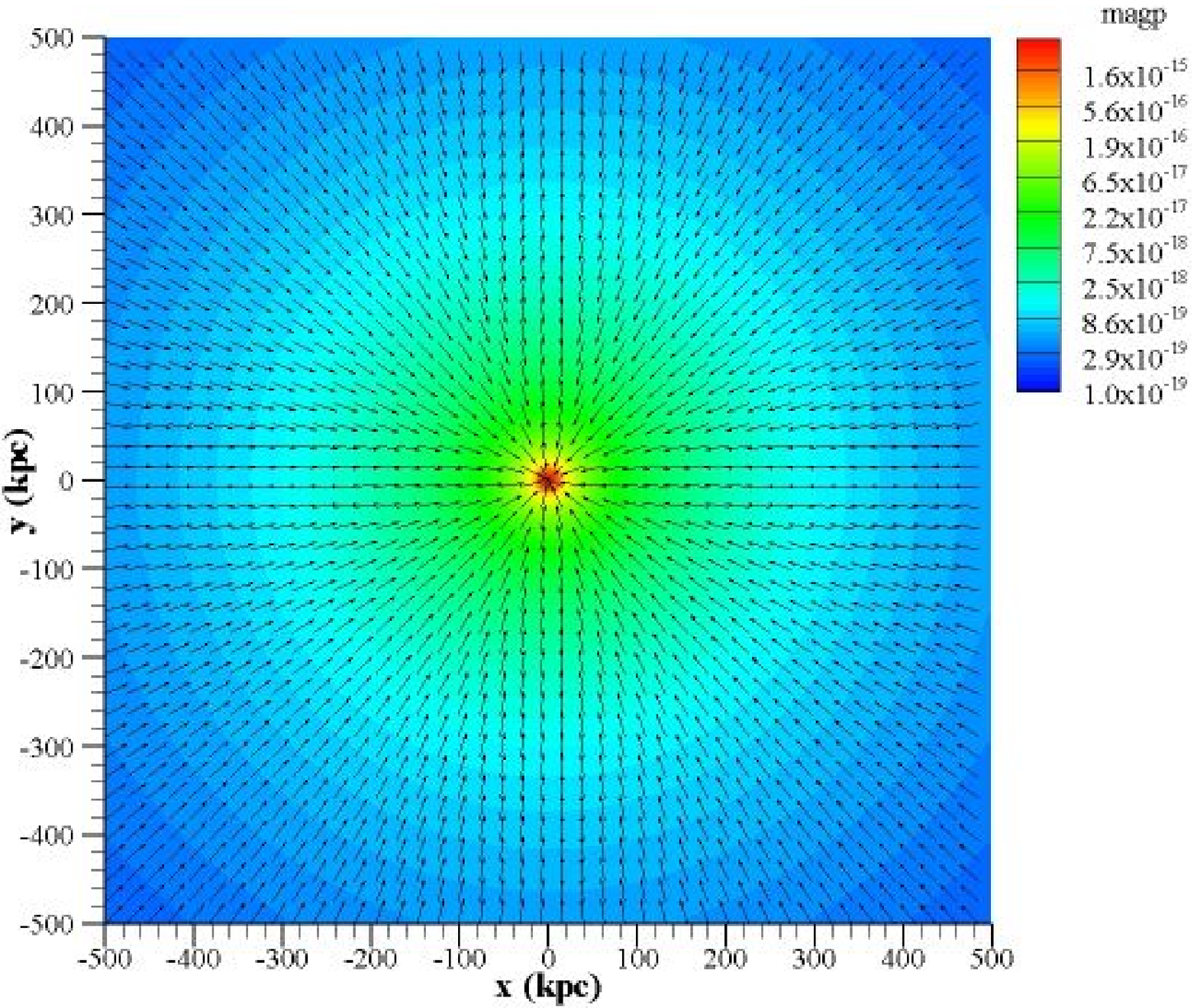}
\includegraphics[trim = 5mm 1mm 7.5mm 5mm, clip, scale=0.5]{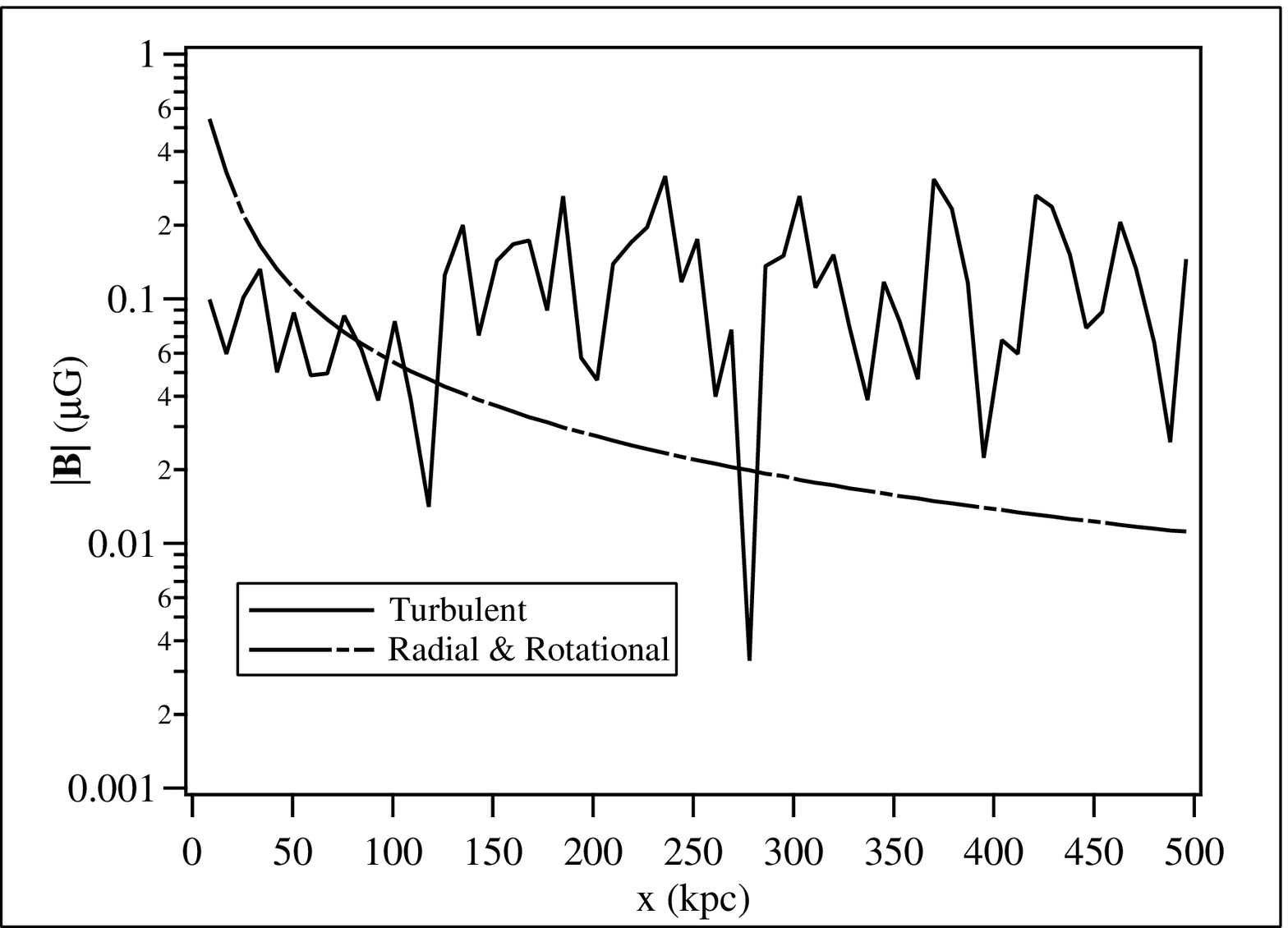}
\includegraphics[trim = 1mm 1mm 1mm 1mm, clip, scale=0.44]{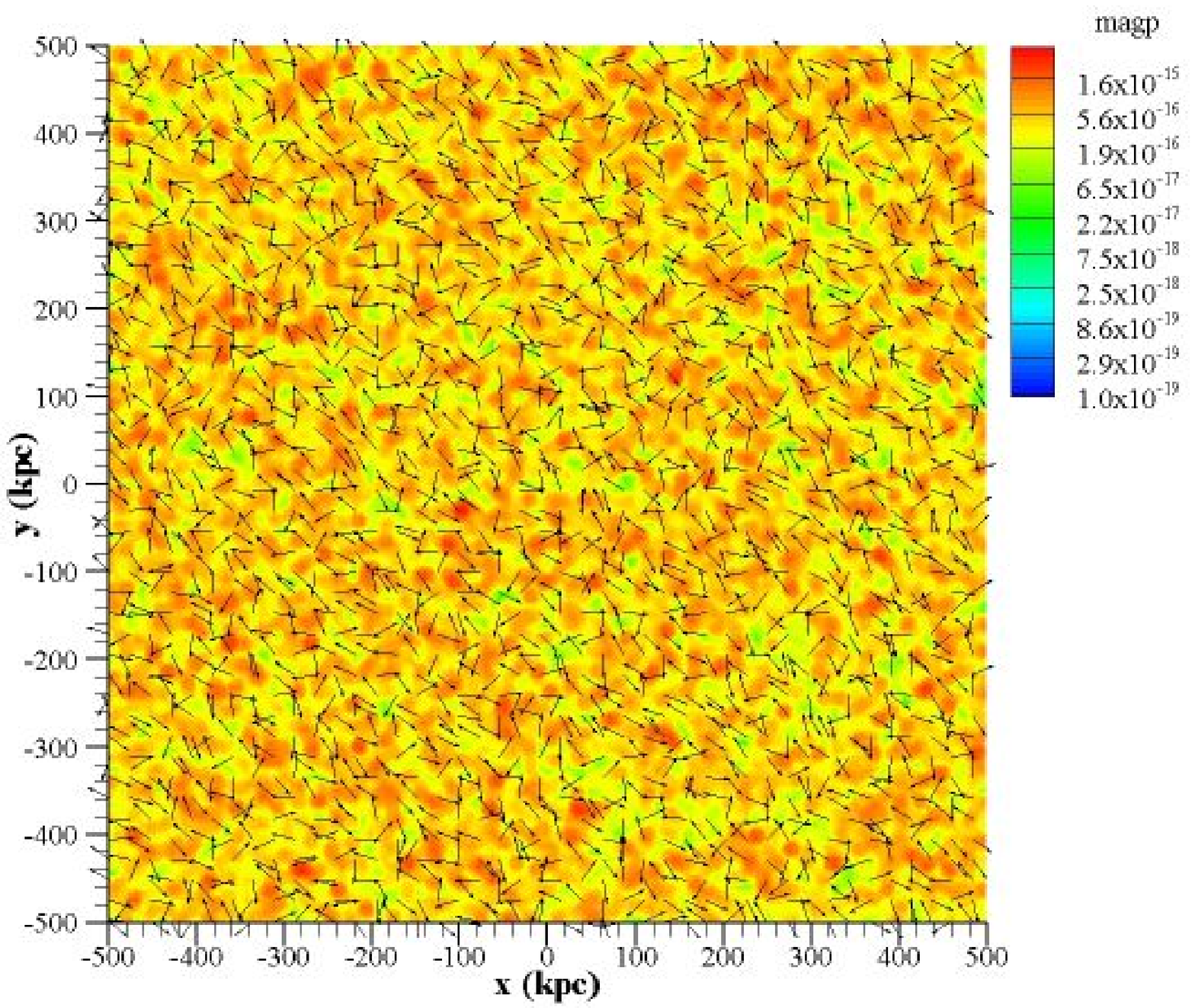}
\caption{Initial magnetic field profiles used in the 2-D planar simulations. The x-y plots show contours of magnetic pressure (in Pa) overlaid by magnetic field unit vectors. The magnitudes of the magnetic field for the three cases used in the simulations - rotational, radial and turbulent - are compared in the lower left plot along x at y=0. In all planar (x,y) simulations the computational mesh was uniform consisting of $128^2$ square cells.}
\label{2DboxInitialProfiles}
\end{figure*}

The turbulent field was constructed based on the approach followed by \citet{2007MNRAS.378..662R} (see also \citealt{2010ApJ...713.1332R}). First, we set up a random field in $\textbf{\textit{k}}$-space, using independent normal random deviates $N_a(0,1)$ (with 0 and 1 being the values of the mean and standard deviation of the random distribution, respectively) for the real and imaginary components:
\begin{eqnarray}
B_{x}(\textbf{\textit{k}})=\left[Re(B_{x}),Im(B_{x})\right]=\left[N_1B,N_2 B\right]\nonumber \\
B_{y}(\textbf{\textit{k}})=\left[Re(B_{y}),Im(B_{y})\right]=\left[N_3B,N_4B\right]
\end{eqnarray}
where the wavenumber magnitude is $|\textbf{\textit{k}}|=({\textit{k}_{x}}^2+{\textit{k}_{y}}^2)^{1/2}$ and subscript ``a'' is a number used to identify different random deviates. \citet{2007MNRAS.378..662R} proposed the following dependence of the magnetic field amplitude on the wave number $|\textbf{\textit{k}}|$:
\begin{eqnarray}
B(k)\propto k^{-11/6}exp \left[-(k/k_0)^4\right] exp\left(-k_1/k\right),
\label{Bspectrum}
\end{eqnarray}
where $k_0$ and $k_1$ control the exponential cutoff terms in the magnetic energy spectra. The former controls the spectra at the large wave numbers and is given by $k_0=2 \pi/\lambda_0$ with $\lambda_0 \propto 43h^{-1}$ and $h$ denoting the normalized Hubble parameter $(0.5-0.8)$. The second, $k_1$, controls the cutoff of the spectra at the small wavelengths; the case we present in this paper uses $k_0=0.095$ and $k_1=0.05$. Once the complex vector field in $\textbf{\textit{k}}$-space has been generated, we perform vector divergence cleaning. Using conventional Einstein tensor summation notation where $\delta_{ij}$ is the Kronecker delta, this operation may be expressed by (e.g. see \citealt{1998ApJS..116..133B}):
\begin{equation}
B_i(\textbf{\textit{k}})= \left(\delta_{ij}-\frac{k_ik_j}{|\textbf{\textit{k}}|^2}\right) B_j(\textbf{\textit{k}}).
\end{equation}
Finally, we perform (complex) inverse Fourier transformation in 2-D to obtain the \textbf{\textit{r}}-space components $[B_{x}(\textbf{\textit{r}}),B_{y}(\textbf{\textit{r}})]$.

Because the precise distribution of magnetic fields in galaxy clusters is uncertain, there is no strong evidence in favor of the assumed spectrum in equation (\ref{Bspectrum}). Therefore, we extend the range of our numerical experiments to include the following two (extreme) cases: a rotational profile and a radial profile as given by equations (\ref{Brotational}) and (\ref{Bradial}), respectively, where $R_0$ is a scaling radius.
\begin{eqnarray}
B_x(x,y)= -B \frac{yR_0}{x^2+y^2},
B_y(x,y)= +B \frac{xR_0}{x^2+y^2}
\label{Brotational}
\end{eqnarray}
\begin{eqnarray}
B_x(x,y)= -B \frac{xR_0}{x^2+y^2},
B_y(x,y)= -B \frac{yR_0}{x^2+y^2}.
\label{Bradial}
\end{eqnarray}
These two additional cases have been used exclusively in the 2-D planar simulations and the results are presented in Section \ref{subsec:PlanarSims}. For both cases the out-of-plane component $B_z$ was set to zero. The initial magnetic field profiles for all three cases are illustrated in Fig. \ref{2DboxInitialProfiles} for a peak magnetic field value of 1 $\mu$G. Hereinafter we shall denote the maximum strength of the magnetic field in our simulations by $B$.
\subsubsection{Boundary conditions}\label{subsec:BConditions}
Boundary conditions in MACH2 are defined by employing ``ghost'' cells and vertices. The simulations in planar geometry were performed using a computational domain that spanned $-L\leq x\leq L$ and $-L\leq y\leq L$ with $L=$500 kpc. At all outer boundaries in this geometry we have imposed a Dirichlet condition for the temperature and the specified value is set equal to the initial value as given in Fig. \ref{InitialProfiles}. Radiative cooling is allowed at these boundaries. For each time step the velocity at the outer boundaries is set to zero (``no slip''), the density is set to the initial value (\ref{InitialProfiles}), and the magnetic field is copied from the real cell adjacent to the boundary to the ghost cell.

For the axisymmetric (r,z) simulations the computational region spans $0\leq r\leq L$ and $0\leq z\leq L$. At the boundaries $(L,z)$ and $(r,L)$ the conditions are the same as those at the outer boundaries of the planar geometry. At $(0,z)$ and $(r,0)$ we impose thermally insulating conditions. Only one case in this computational domain employs a magnetic field and it is kept frozen everywhere throughout the simulation.
\section{numerical experiments}\label{sec:NumExperiments}
A series of numerical experiments were performed in 2-D axisymmetric (r,z) and 2-D planar (x,y) geometry with the two initial plasma profiles described in Section \ref{subsec:IConditions}, LCD and HCD, and three initial magnetic field profiles described in Section \ref{subsec:IBConditions} for a wide range of combinations and simulation physics. Hereinafter, it will be implied that when we present simulation results without a magnetic field, thermal conduction has been computed isotropically, whereas for simulations that include a magnetic field thermal conduction has taken into account the anisotropic effects (see equations \ref{eqn:thcondtensor} and \ref{eqn:heatflux}).

Table \ref{tab:2DAxisymmetricSims} lists the cases carried out in 2-D axisymmetric geometry. A goal of the simulations in this geometry was to assess the sensitivity of the global heat flux factor \textit{f} on the plasma density near the A2199 core under idealized simulation physics. Specifically, no hydrodynamics were permitted in any of these simulations and the magnetic field, when used, was held frozen. Table \ref{tab:2DPlanarSims} outlines our cases in 2-D planar geometry. In all of these cases no control of the thermal conductivity was exercised and the anisotropic heat flux was computed self-consistently. The focus of these simulations was the early dynamics in the near-core region of the cluster and the results are presented in Section \ref{subsec:PlanarSims}.
\begin{table}
\caption{Numerical experiments in 2-D axisymmetric (r,z) geometry. Only thermal conduction and radiation were included in the simulation physics (hydrodynamics were not allowed in these cases). The heat flux factor \textit{f} was either specified or allowed to evolve freely.} % title name of the table
\centering % centering table
\begin{tabular}{ccc}
\hline
\multicolumn{2}{c}{Initial Conditions} & Heat Flux\\
\cline{1-2}
$n_e$ \& $T_e$ & $B (\mu G)$ [Profile] & Factor, \textit{f} \\
\hline
LCD   & 0.0   [N/A] & 0.4 \\
LCD   & 0.0   [N/A] & 0.5 \\
LCD   & 0.0   [N/A]  & 0.6 \\
LCD   & 0.0   [N/A] & 1.0 \\
LCD   & $1.0\times10^{-12}$   [turbulent \& frozen]  & free \\
LCD   & 0.1   [turbulent \& frozen]  & free \\
LCD   & 1.0   [turbulent \& frozen]  & free \\
HCD   & 0.0   [N/A] & 0.5 \\
HCD   & 0.0   [N/A]  & 0.6 \\
\hline
\end{tabular}
\label{tab:2DAxisymmetricSims}
\end{table}
\begin{table}
\caption{Numerical experiments in 2-D planar (x,y) geometry. The heat flux factor \textit{f} was allowed to evolve freely. (Simulation physics nomenclature: ``C''= Thermal conduction, ``R''= Radiation cooling, ``H''= Hydrodynamics, ``M''= Magnetohydrodynamics.)} % title name of the table
\centering % centering table
\begin{tabular}{ccc}
\hline
\multicolumn{2}{c}{Initial Conditions} & Simulation\\
\cline{1-2}
$n_e$ \& $T_e$ & $B (\mu G)$ [Profile] & Physics\\
\hline
LCD   & 0.0   [N/A] & CR\\
LCD   & 0.0   [N/A] & CRH\\
LCD   & 1.0   [turbulent \& frozen] & CR\\
LCD   & 1.0   [rotational] & CRM\\
LCD   & 1.0   [radial] & CRM\\
LCD   & 1.0   [turbulent] & CRM\\
LCD   & $1.0\times10^{-3}$   [turbulent] & CRM\\
HCD   & 1.0   [turbulent] & CRM\\
HCD   & $1.0\times10^{-3}$   [turbulent] & CRM\\
\hline
\end{tabular}
\label{tab:2DPlanarSims}
\end{table}
\subsection{2-D simulations in axisymmetric (r,z) geometry}\label{subsec:AxisymSims}
The main intent of the simulations in 2-D axisymmetric geometry was to conduct a preliminary assessment of the sensitivity of the ICM solution to the prescribed initial conditions before invoking full MHD simulations. Due to the dependence of the cooling rate on the square of the electron density, $\Phi_{eR}\propto n_e^2 \sqrt{T_e}$, we expect the solution will depend strongly on the near-core gas density (e.g. see \citealt{2008ApJ...681..151C}). In fact, if one assumes that the dominant competing mechanisms establishing the thermal balance in the ICM are thermal conduction and radiation only, and all other mechanisms are captured by a single effective reduction of the thermal conductivity (represented by the quantity \textit{f}), we can estimate of this sensitivity from the steady-state energy equation \ref{ICMenergyIdeal}, by computing directly the heat flux factor $\textit{f}$ of the cluster. The value of the ICM density in the cluster core ($R\approx$0.95 kpc), based on the \textit{Chandra} observations, is $n_e\approx 0.115$ $cm^{-3}$, whereas the LCD model at the same location gives $n_e\approx 0.075 cm^{-3}$. Since the temperature predicted by the LCD and HCD models is approximately the same at this location, we use the LCD model to estimate $\nabla \cdot \left(\kappa_{sp} \nabla T_e\right)$. By approximating the core region as a small finite cylindrical volume with dimensions $\Delta r=\Delta z\approx$ L/128=3.906 kpc we find:
\begin{enumerate}
  \item $\textit{f}\approx 0.8$ when $n_e\approx 0.115$ $cm^{-3}$ 
  \item $\textit{f}\approx 0.5$ when $n_e\approx 0.075$ $cm^{-3}$,
\end{enumerate}
which illustrates to first order the sensitivity of idealized solutions on the near-core plasma density. 

The significance of this sensitivity may be better appreciated by comparing the following relevant time scales. First, the characteristic time $\tau_{q}$ associated with the isothermalization of a region with a temperature gradient $\sim\Delta T_e/\Delta R$, may be estimated by setting
%(e.g. see \citealt{2008ApJ...681..151C} or \citealt{2007ApJ...663..816A})
\begin{equation}
\frac{3}{2}n_e k_B \frac{\partial T_e} {\partial t}=-\nabla \cdot \bmath{q}
\end{equation}
yielding
\begin{equation}
\tau_q \approx \frac{3}{2}n_e k_B \frac{\Delta R^2} {\textit{f}\Delta \kappa_{sp}}\propto \frac{n_e \ell^2}{\textit{f} T_e^{5/2}}
\label{eqn:CondTime}
\end{equation}
(with $\ell \sim\Delta R$). Next, the characteristic time $\tau_{\Phi}$ in which the plasma radiates away all of its internal energy is obtained in a similar manner yielding 
%(e.g. see \citealt{2008ApJ...681..151C}),
\begin{equation}
\tau_{\Phi} \approx \frac{3}{2} \frac{n_e k_B T_e}{a c_{\ell}\rho\chi_p {T_e}^4} \propto \frac{T_e^{1/2}}{n_e}.
\end{equation}

If the mechanism(s) in the ICM produce an effective reduction of the thermal conductivity $\textit{f}\approx 0.5$, then the conduction time would be more than double the radiation time, assuming the plasma density near the core is as high as the observed value. In other words, without any additional heating source(s), these estimates suggest that thermal conduction from the outer regions of the cluster cannot avert runaway cooling near the core.
\begin{figure}
\includegraphics[trim = 1.5mm 1mm 7.5mm 1mm, clip, scale=0.47]{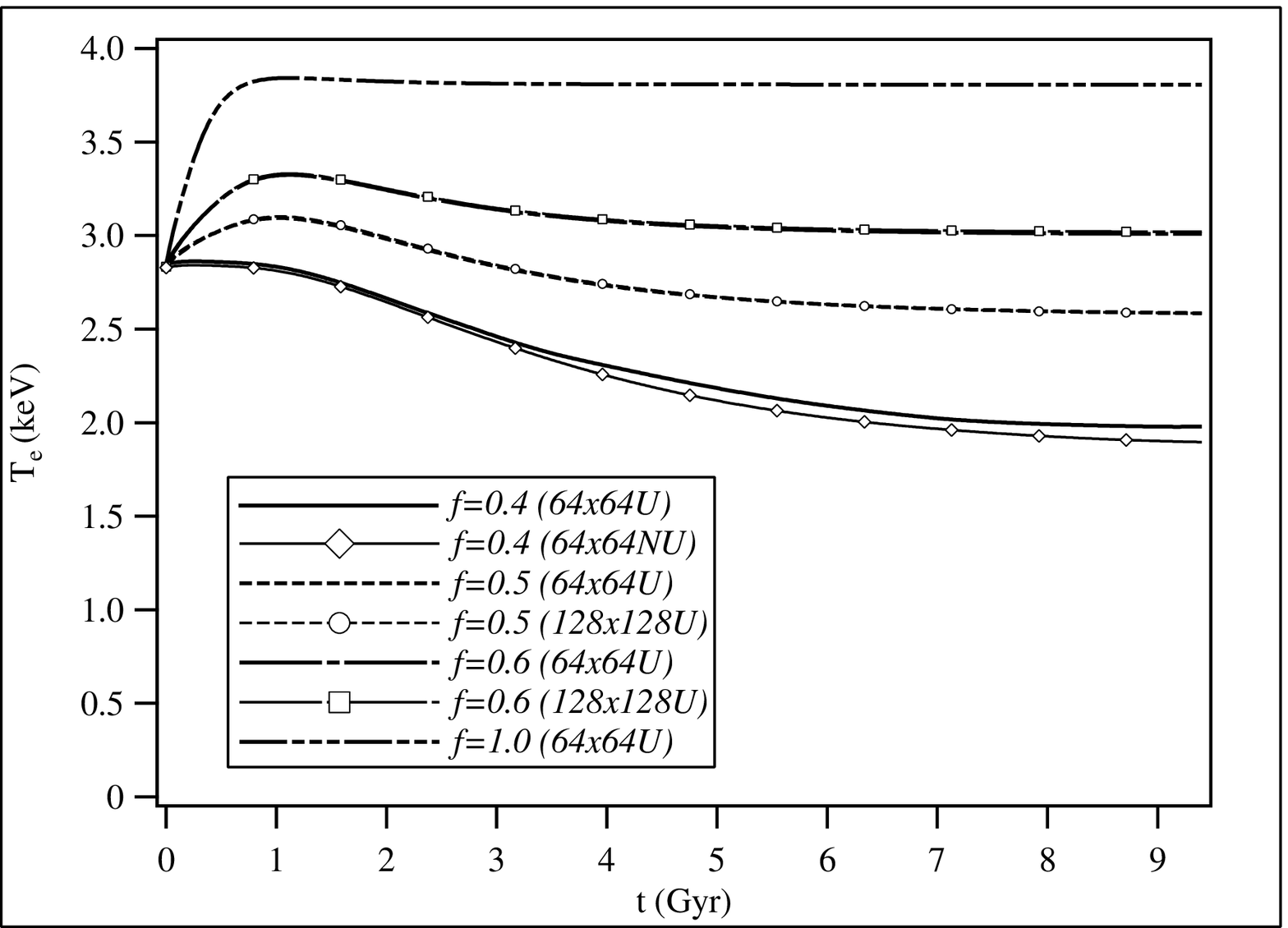}
\includegraphics[trim = 1.5mm 1mm 7.5mm 1mm, clip, scale=0.47]{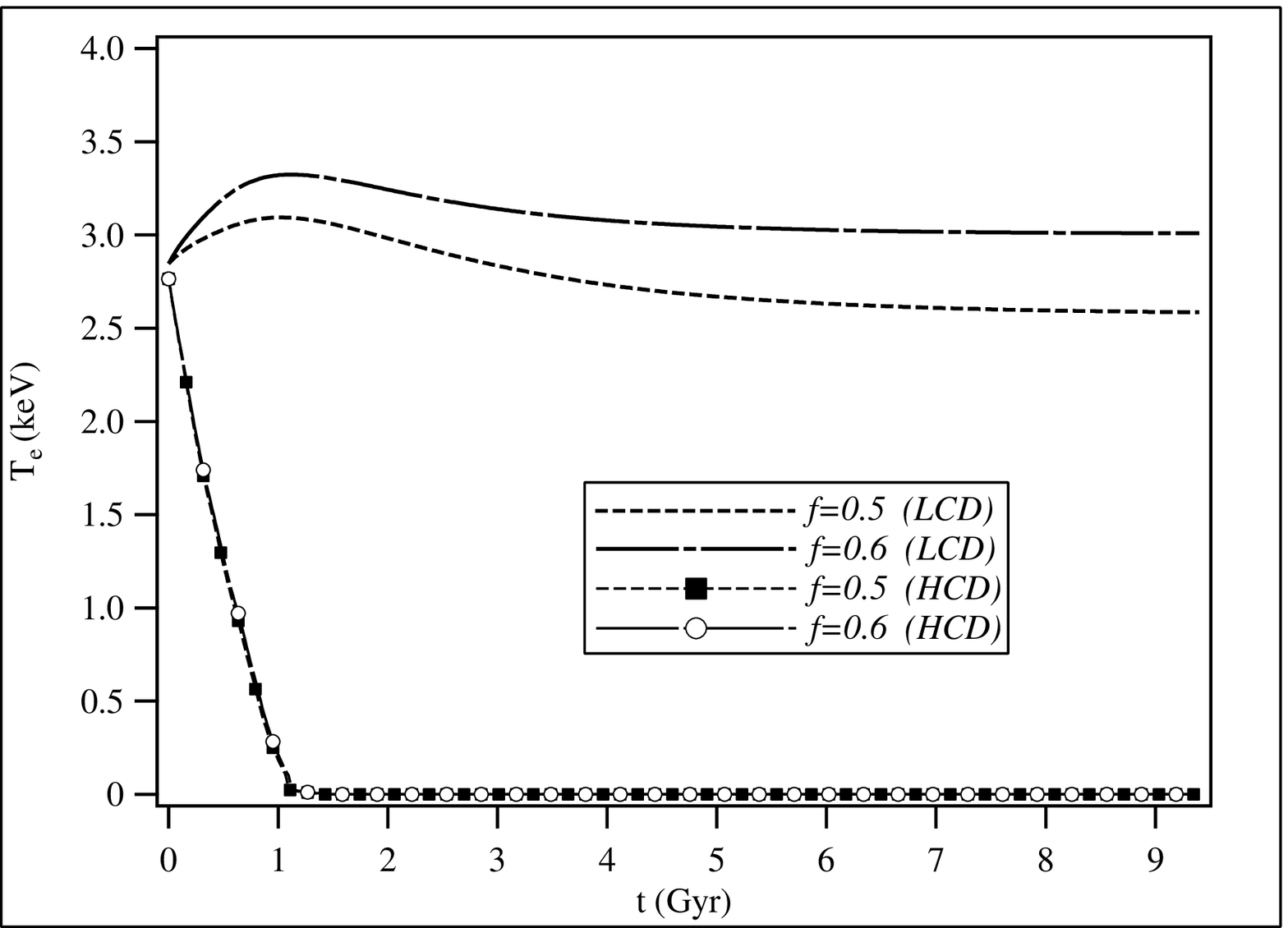}
\caption{Results from 2-D axisymmetric simulations with thermal conduction and radiation only. The solutions are plotted for different values of the heat flux factor \textit{f}, at $R=$20 kpc. Top: Results using the LCD initial conditions for $64^2$ and $128^2$ uniform (U) computational meshes. The $64^2$ non-uniform (NU) mesh used quadratic spacing with highest resolution of 0.25 kpc at the cluster core. Bottom: Comparisons of solutions for the LCD and HCD initial conditions at $\textit{f}$=0.5 and 0.6 (it is noted that differences in the two HCD solutions are negligible).}
\label{RZsolutions}
\end{figure}
\begin{figure}
\includegraphics[trim = 1mm 1mm 1mm 1mm, clip, scale=0.61]{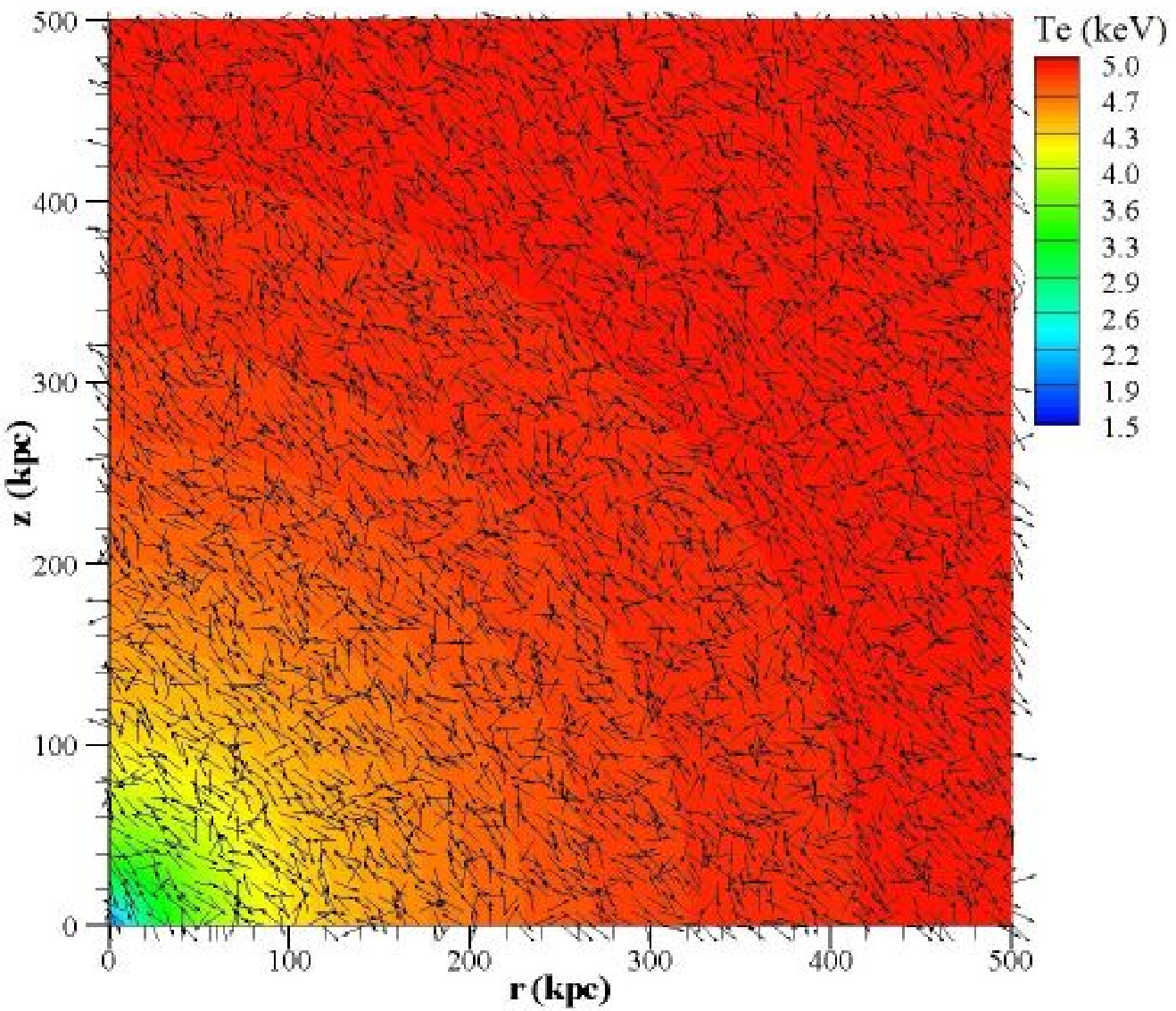}
\includegraphics[trim = 1.5mm 1mm 5mm 1mm, clip, scale=0.47]{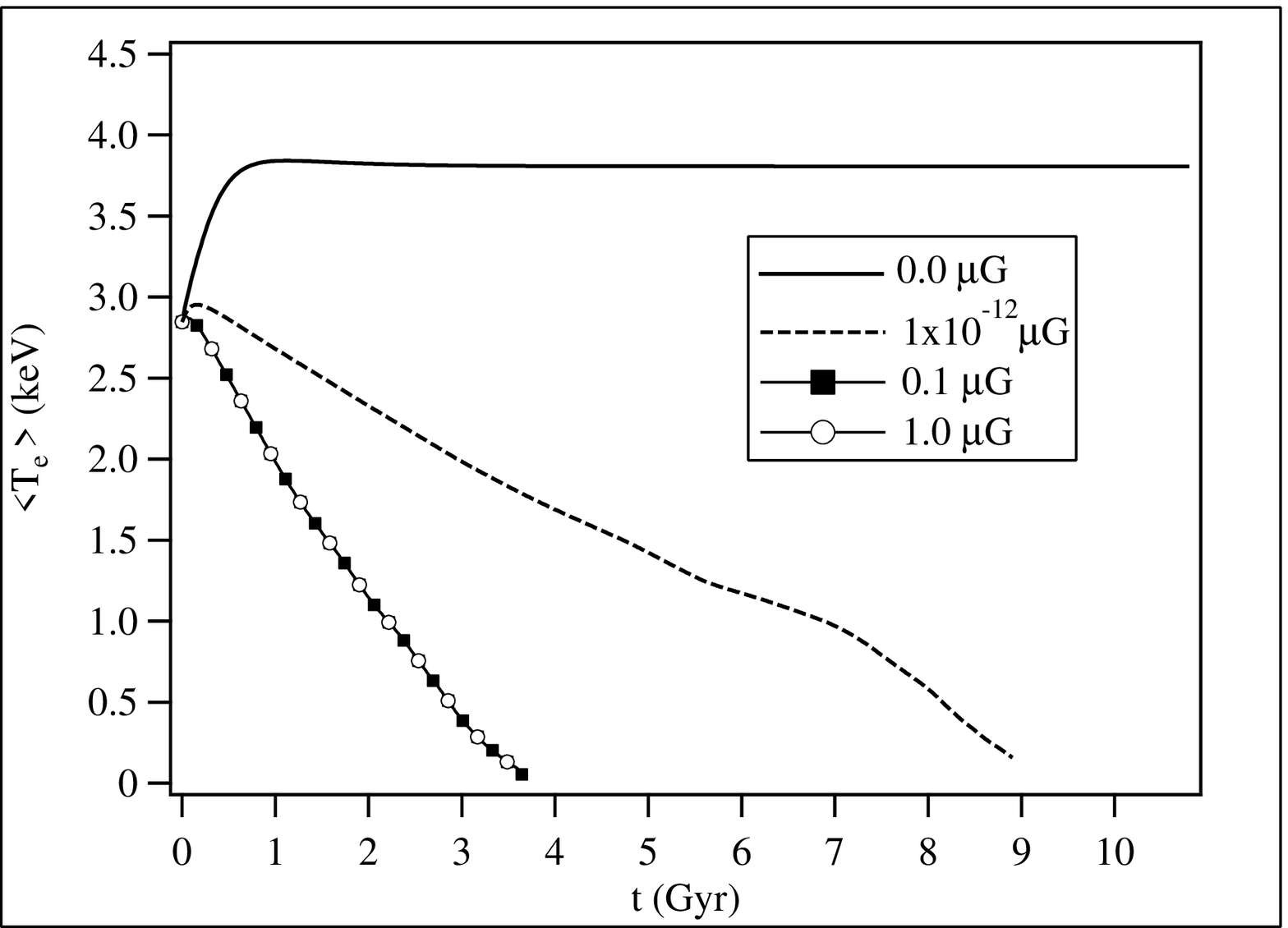}
 \caption{Results from 2-D axisymmetric simulations with anisotropic thermal conduction, radiation, and a turbulent magnetic field. The magnetic field was frozen throughout the simulation. Top: Temperature contours overlaid by magnetic field unit vectors at t=0. Bottom: Evolution of the average temperature at $R=$20 kpc for different values of the peak magnetic field amplitude allowed. It is noted that differences between the solutions for the 0.1 and 1 $\mu$G are negligible.}
\label{RZsolutionsFrozenField}
\end{figure}

Indeed, 2-D axisymmetric simulations with MACH2 that incorporated only thermal conduction and radiation (MHD was turned off in this series of calculations) show that thermal balance is achieved at a relatively low value of $\textit{f}$ (between 0.5-0.6) for the LCD profile, whereas the HCD profile leads to the collapse of the core for the same values of $\textit{f}$. Figure \ref{RZsolutions} (top) shows the evolution of the computed temperature at $R=$20 kpc for different values of $\textit{f}$, when LCD initial profiles were used. Mesh sensitivity calculations confirmed that a $64^2$ uniform mesh provided sufficient resolution. Figure \ref{RZsolutions} (bottom) compares the LCD solutions to those obtained using the HCD profiles for $\textit{f}=$ 0.5 and 0.6, showing collapse of the core in less than 1 Gyr. Thus, determining a global value for \textit{f} that would sustain thermal balance in the cluster may be in general instructive; quantitatively, however, it is of less significance due to the sensitivity of \textit{f} on the central density.

Due to the extremely high Hall parameters in the ICM ($\Omega_e>10^9$ in A2199) the potential impact of turbulent magnetic fields in the transport of heat from the outer (hot) regions of the cluster has been widely recognized. Although the evolution and impact of MHD instabilities, specifically of the HBI (e.g. see \citealt{2008ApJ...673..758Q}; \citealt{2008ApJ...677L...9P}) and of the magneto-thermal instability (MTI) (e.g. see \citealt{2000ApJ...534..420B}; \citealt{2008ApJ...688..905P}), remain an ongoing topic of study, we performed here an idealized numerical experiment with a turbulent initial magnetic field (see Section \ref{subsec:IBConditions}) that was kept frozen throughout the simulation. The intent was to emulate a scenario in which continuous galaxy motions sustain a nearly-isotropic turbulent field. The simulation case was motivated in part by the work of \citet{2010ApJ...713.1332R}, who suggest it is possible that turbulent velocity fields in the cluster may drive (or stir) turbulence in the magnetic field in a manner that allows for sufficiently rapid transport of heat from the outer regions of the ICM to sustain thermal balance.

Figure \ref{RZsolutionsFrozenField} (top) shows contours of the initial temperature overlaid by unit vectors of the initial magnetic field as implemented in our 2-D axisymmetric simulation. For this case the peak magnitude of the magnetic field did not exceed 1 $\mu$G. The evolution of the computed temperature at $R=R_c$ is shown in Fig. \ref{RZsolutionsFrozenField} (bottom). Also shown for comparison is the case with no magnetic field. For all cases in this series of simulations the thermal conductivity was computed self-consistently, accounting fully for the anisotropy in the heat flux, but magnetohydrodynamics were excluded. We found that thermal conduction alone is unable to avert a cooling catastrophe, even when the magnetic field is completely tangled. The collapse of the core for the 1-$\mu$G case was found to be insensitive to the value of peak magnetic field, which is expected due to the high values of $\Omega_e$. In this arrangement a change in the collapse rate was obtained only when the peak magnetic field magnitude was lowered by several orders of magnitude (where $\Omega_e\rightarrow 1$). 

By comparison, \citet{2010ApJ...713.1332R} found in their 3-D simulations of A2199 that a cooling catastrophe could be averted. The authors applied a driving force that resulted not only in keeping the magnetic field tangled but also in turbulent flows that introduced additional physical effects such as mixing, which we do not account for in our 2-D axisymmetric simulation. We also recognize that the real topology of the magnetic field is unknown and may differ significantly from the idealized model described in Section \ref{subsec:IBConditions}. Nevertheless, in view of the sensitivity of thermal balance calculations on the plasma density it appears that hypotheses about ``stirring'' or other dynamical activity in the ICM, proposed as mechanisms that could sustain thermal balance in the cluster, would only be strengthened if such mechanisms sustained balance in the cluster for a wide range of possible core densities. \citet{2010ApJ...713.1332R} employed an initial temperature profile that was in close agreement with the \textit{Chandra} data and a plasma density profile with values almost six times lower than the observed data near the cluster core.
\subsection{2-D simulations in planar (x,y) geometry and in hydrostatic non-equilibrium}\label{subsec:PlanarSims}
In this section we perform an idealized study in two dimensions near the core of A2199, focusing on possible mechanisms that may overcome HBI-driven collapse (\citealt{2009ApJ...703...96P, 2010ApJ...713.1332R}). In the 3-D MHD simulation cases of \citet{2009ApJ...703...96P} and \citet{2010ApJ...713.1332R} some initial ICM states did not achieve thermal balance but all cases imposed hydrostatic equilibrium at t=0. Here, we deliberately violate this condition only as a means to generate and study near-core flow dynamics ($<$20 kpc) that may possibly be associated with AGN activity. 

We deviate from hydrostatic equilibrium by prescribing the spherically-symmetric initial conditions LCD and HCD (Fig. \ref{InitialProfiles}) in a 2-D planar geometry. We note that although both these profiles correspond to hydrostatic equilibrium in three dimensions, they do not preserve such equilibria in two dimensions, thereby producing subsonic wave motion in the plasma. Here we focus only on the early dynamics of the MHD flow field, $t<$ 2.5 Gyr, near the cluster center ($R<$20 kpc), since several simulations of A2199 have predicted a cooling catastrophe due to the HBI within this time for a wide range of magnetic field strengths and profiles. Assuming a weak magnetic field initially in the direction of the gravitational force, the HBI has a characteristic growth rate (\citealt{2008ApJ...673..758Q})
\begin{equation}
\abs{\gamma_{HBI}}\backsimeq \left(g\frac{\nabla_R T_e}{T_e}\right)^{1/2}
\label{eqn:HBIGrowth}
\end{equation}
$\approx$ (0.125 Gyr)$^{-1}$. Thus the instability is allowed approximately 20 growth times in $\sim$2.5 Gyr.

\begin{figure}
\includegraphics[trim = 1.5mm 1mm 5mm 2mm, clip, scale=0.50]{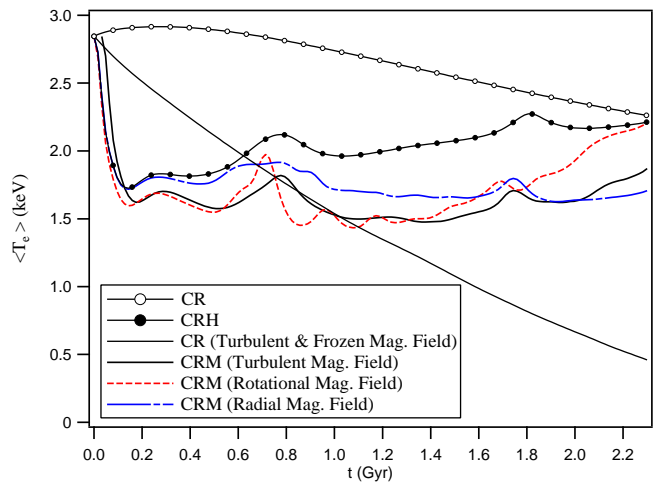}
 \caption{Comparisons of the average temperature at $R=$20 kpc, computed in planar geometry for cases with the same (LCD) profiles but different simulation physics (see Table \ref{tab:2DPlanarSims} for simulation physics nomenclature). For all cases with a magnetic field the maximum strength was 1 $\mu$G.}
\label{AllZNcases2}
\end{figure}

We find in the 2-D planar simulations with a turbulent initial magnetic field (see Fig. \ref{2DboxInitialProfiles} bottom right) that the average temperature at $R=$20 kpc is sustained and a cooling catastrophe does not occur, as shown in Fig. \ref{AllZNcases2}. For comparison the solutions with the same simulation physics are also shown in Fig. \ref{AllZNcases2} for the following cases: (1) radial and (2) rotational initial magnetic fields, (3) turbulent initial magnetic field profile but held frozen throughout the simulation (a case with no MHD but one that accounted for anisotropic thermal conduction and radiative cooling), (4) isotropic thermal conduction and radiation without MHD, and (4) isotropic thermal conduction, radiation and hydrodynamics but no magnetic field. Similar to the cases of isotropic conduction no catastrophic cooling is found in the simulations with large-scale flows (induced by starting out of hydrostatic equilibrium), independent of the initial configuration of the magnetic field and despite a significant reduction of the heat flux factor.
%%%%%%%%%%%%%%%%%%%%%%%%%%%%%%%%%%%%%%%%%%%
%%%%%%%%%%%%%%%%%%%%%%%%%%%%%%%%%%%%%%%%%%%
\begin{figure*}
\includegraphics[trim = 1.5mm 1mm 5mm 2mm, clip, scale=0.47]{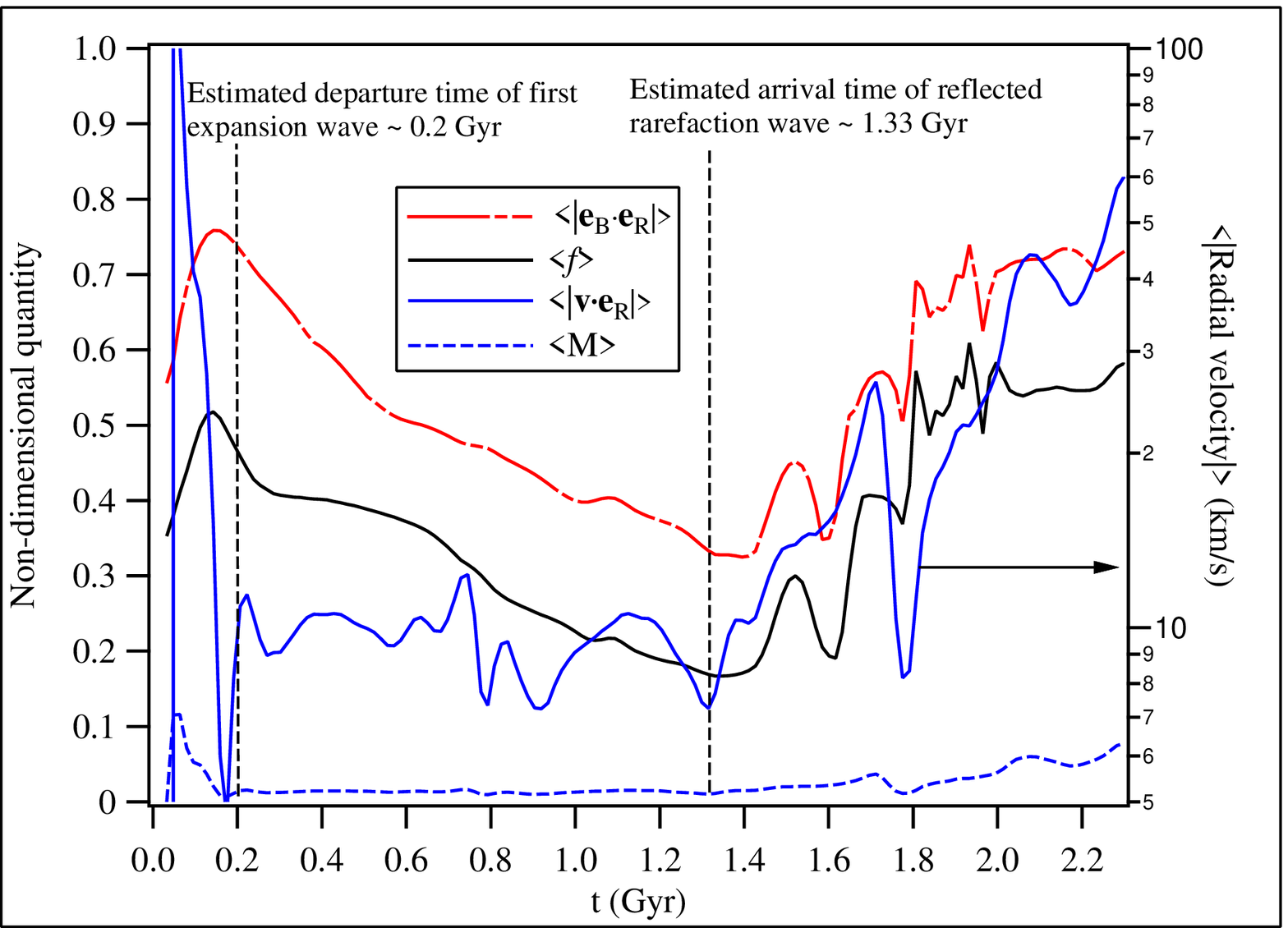}
\includegraphics[trim = 1mm 1mm 24mm 1mm, clip, scale=0.5]{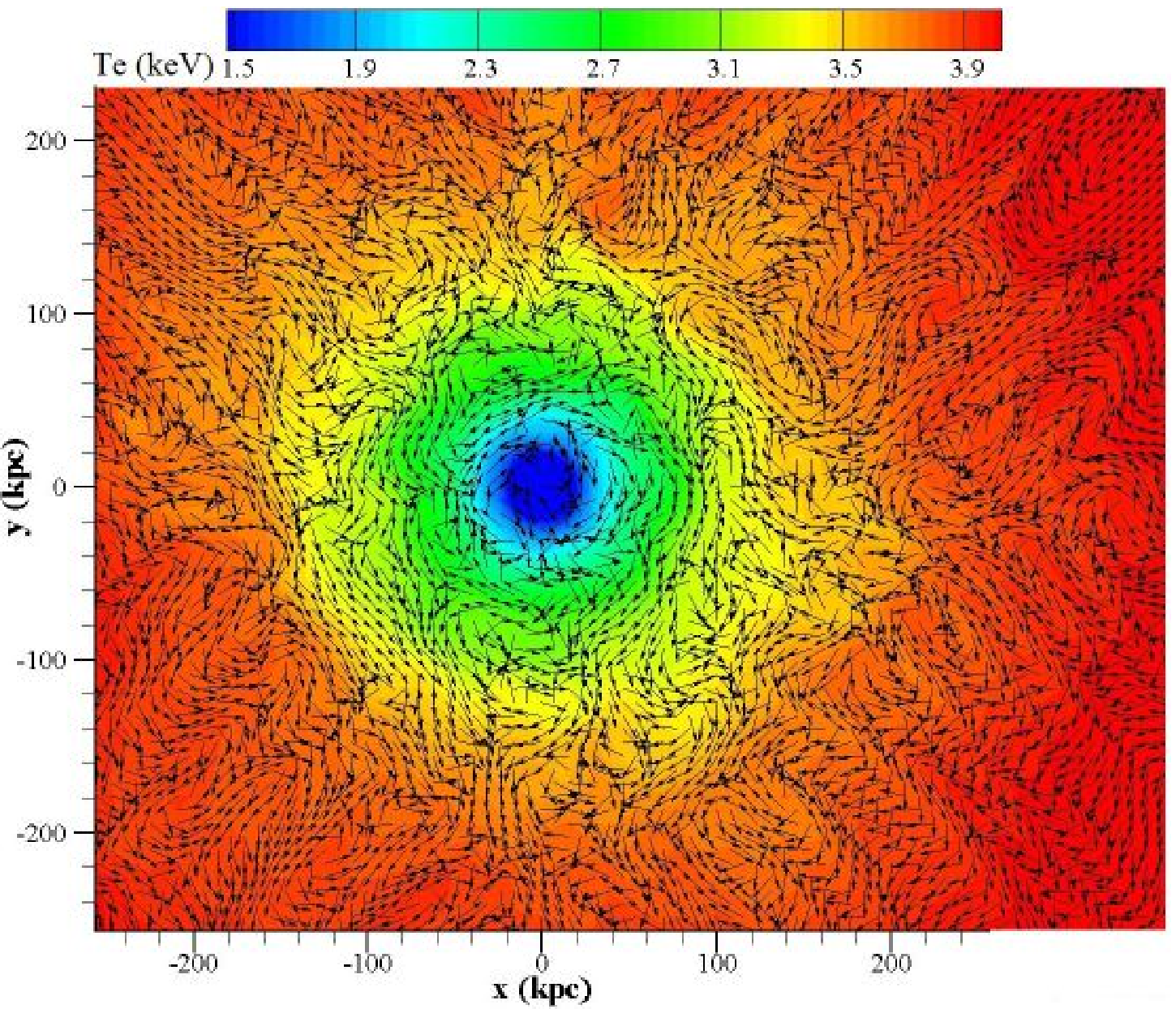}
 \caption{Results from 2-D planar simulations with CRM simulation physics starting with the LCD initial condition for the density and temperature and a turbulent initial magnetic field with maximum magnitude of 1 $\mu$G. Left: Evolution of the cosine of the angle between the magnetic field and radial position vectors, $\langle\left| \bmath{\hat{e}}_B \cdot \bmath{\hat{e}}_R\right|\rangle$, heat flux factor $\langle\textit{f}\rangle$, absolute radial velocity $\langle\left|\textbf{\textit{v}} \cdot \bmath{\hat{e}}_R\right|\rangle$ and Mach number $\langle\left|\textbf{\textit{v}} \cdot \bmath{\hat{e}}_R\right|\rangle/\langle c_s\rangle$, averaged over the area of a disc of radius $R$=20 kpc ($\pm \Delta$),  Right: Temperature contours (in eV) overlaid by magnetic field unit vectors at t=1.27 Gyr showing the orientation of the field to be largely rotational near the cluster core.}
\label{fs-LCD}
\end{figure*}
%%%%%%%%%%%%%%%%%%%%%%%%%%%%%%%%%%%%%%%%%%
Figure \ref{fs-LCD} (left) plots the evolution of the cosine of the angle between the magnetic field and radial position vectors, $\langle\left|\bmath{\hat{e}}_B \cdot \bmath{\hat{e}}_R\right|\rangle$ and the heat flux factor $\langle\textit{f}\rangle$ in the case of the turbulent initial magnetic field and the LCD profile. Unless otherwise noted all quantities have been averaged over the area of a ring of radius 20 kpc and maximum thickness of $\Delta$, which denotes the side length of a single computational cell (recall our mesh consists of square cells). Also plotted in Fig. \ref{fs-LCD} (left) are the absolute radial velocity $\langle|\textbf{\textit{v}} \cdot \bmath{\hat{e}}_R|\rangle$ and Mach number $\langle|\textbf{\textit{v}} \cdot \bmath{\hat{e}}_R|\rangle/\langle c_s\rangle$ where $c_s$ is the adiabatic acoustic speed. 
\begin{figure*}
\includegraphics[trim = 1mm 1mm 1mm 1mm, clip, scale=0.33]{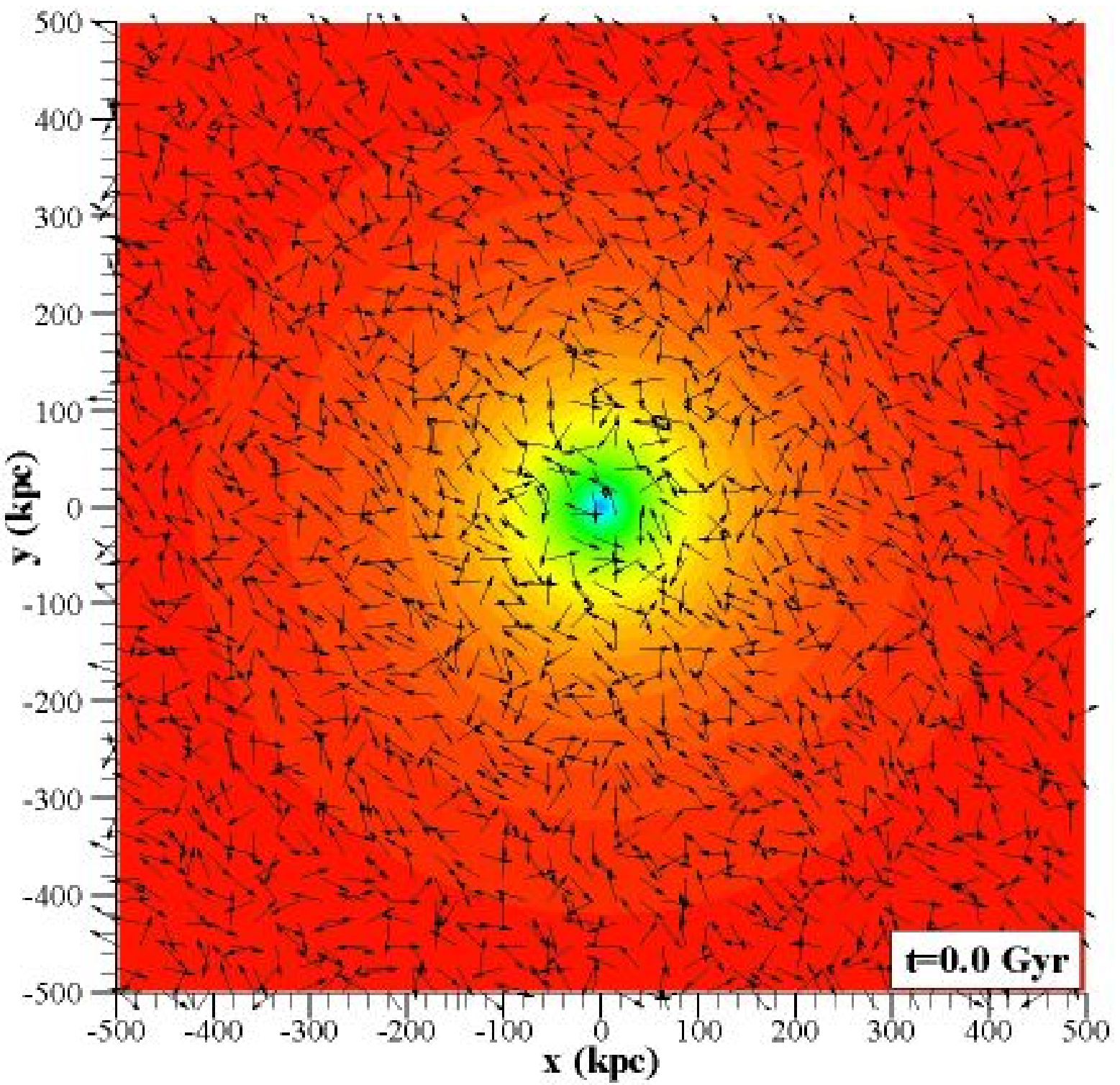}
\includegraphics[trim = 1mm 1mm 1mm 1mm, clip, scale=0.28]{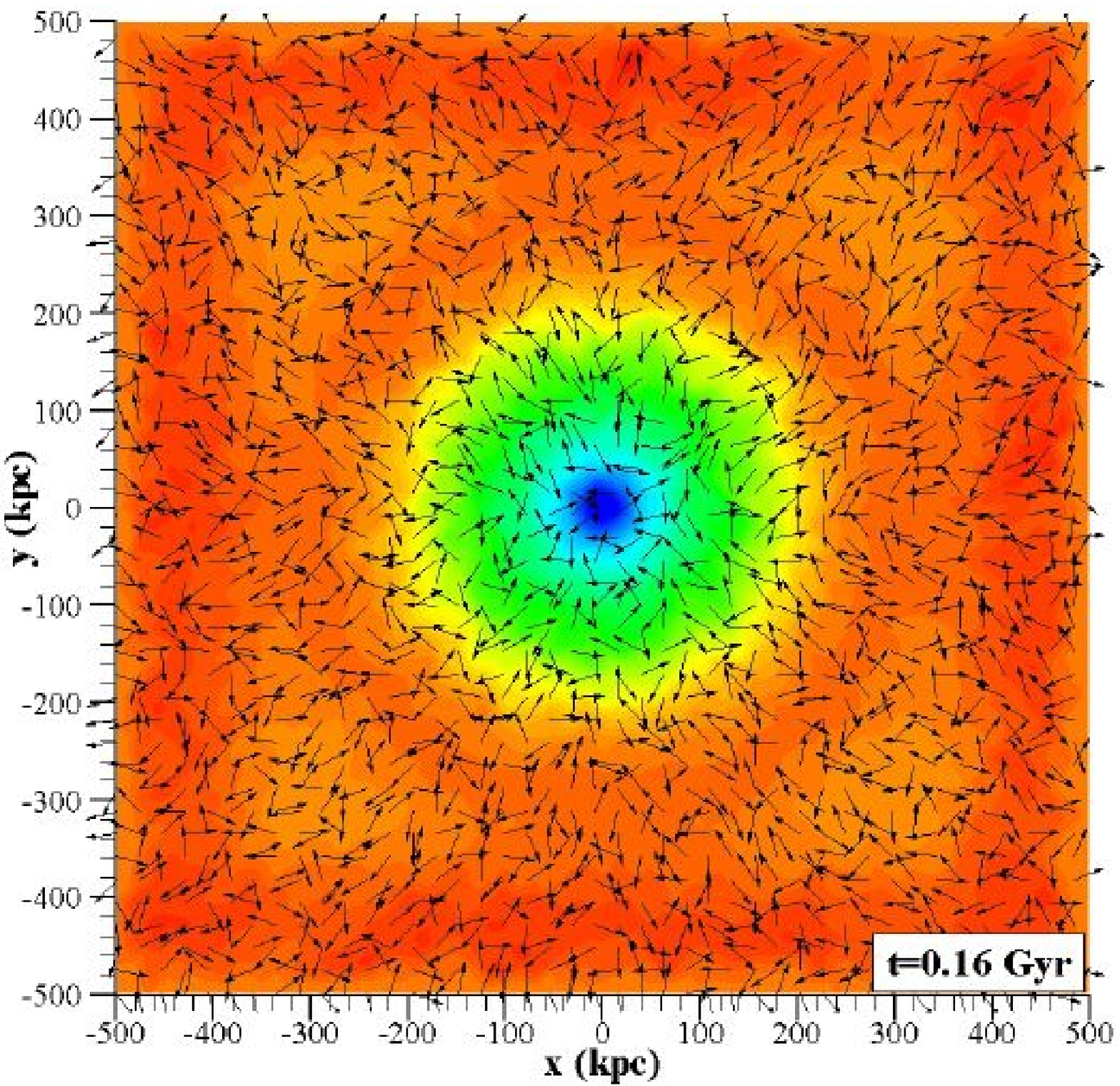}
\includegraphics[trim = 1mm 1mm 1mm 1mm, clip, scale=0.44]{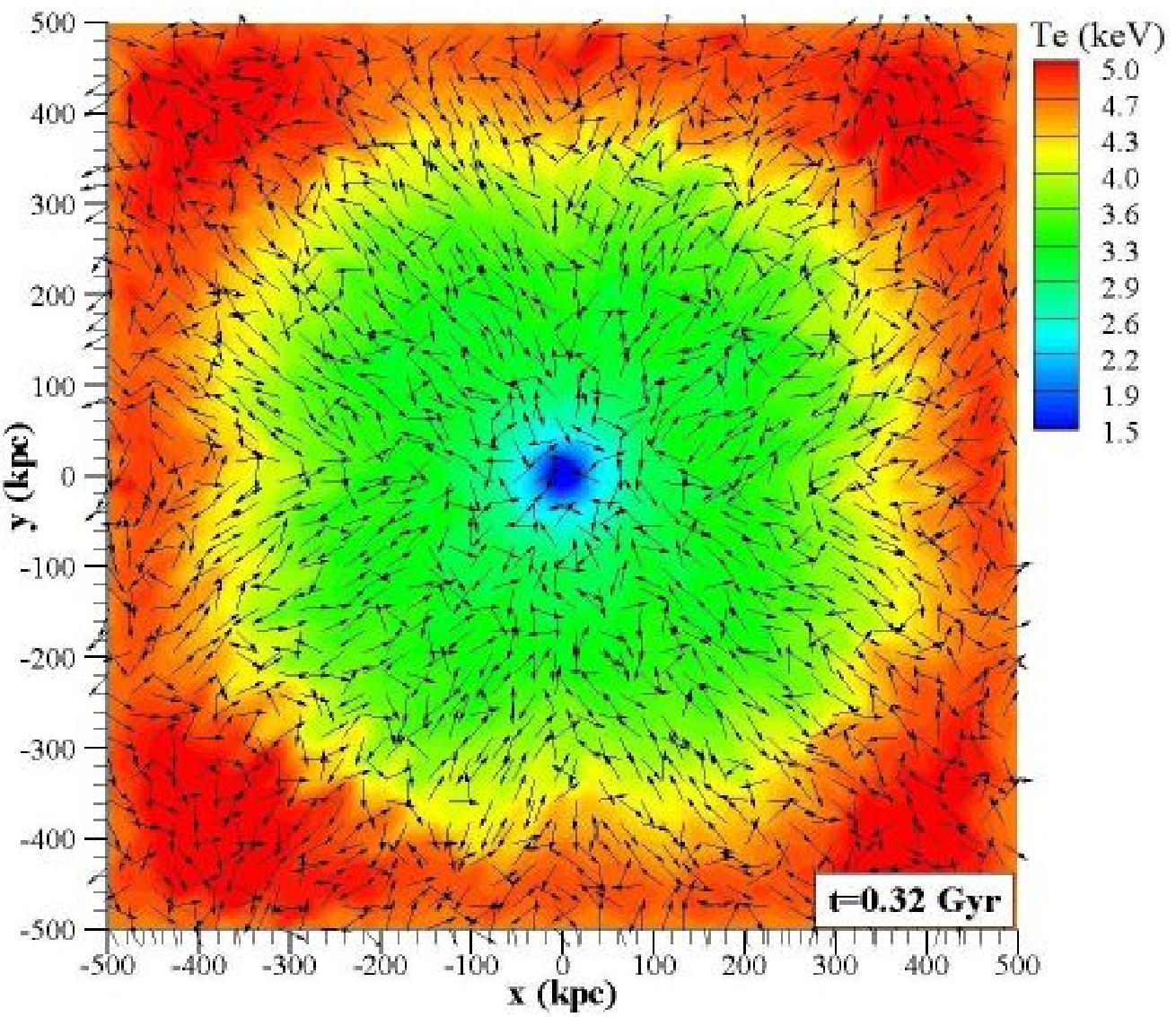}
\caption{Evolution of the temperature and magnetic (unit vector) field for the case of the LCD initial profile and a 1 $\mu G$ turbulent initial magnetic field. The calculations employed CRM simulation physics in planar geometry.}
\label{2DLCD}
\end{figure*}
%%%%%%%%%%%%%%%%%%%%%%%%%%%%%%%%%%%%%%%%%%%

The evolution of the velocity field as shown in Fig. \ref{fs-LCD} (left) defines a characteristic interval of time, $0.2<t<1.3$ Gyr. The interval is bounded by two characteristic times. The shortest time is associated with the hydrostatic imbalance that is imposed initially in the system, producing a subsonic expansion with a characteristic speed of $\sim$100 km/sec ($\langle M\rangle\sim$0.1). The time it takes for the expansion wave to reach a radius of 20 kpc is $\sim$20 kpc/100 km/sec=0.2 Gyr. As shown in Fig. \ref{fs-LCD} (left), upon passage of this first wave the rarefaction flow at $R=$20 kpc moves at an approximately constant speed, $\sim$10 km/s ($\langle M \rangle\sim$0.01) for about another 1.1 Gyr. Such ``winds'' are, in fact, observed in luminous galaxies (and are commonly labeled broad-line regions, e.g. see \citealt{2010arXiv1004.2923O}). In the ensuing discussions we focus in this ``quiet'' rarefaction period $0.2<t<1.3$ Gyr as an emulator of low-speed flow activity (possibly associated with an AGN). Figure \ref{2DLCD} plots 2-D temperature contours and unit vectors of the magnetic field at t=0, 0.16, and 0.32 Gyr showing the evolution of the expansion.

The two-way transit time for sound waves generated near the core is $\sim$ $2L/c_s$=1.3 Gyr. At about this time the heat flux factor and cosine of the magnetic field angle at $R=$20 kpc have reached their minimum values of 0.17 and 0.35 (angle=70 deg), respectively (note that the angle in fact decreases within $t\approx$0.2 Gyr to about $\sim$45 deg from the initial value of $\sim$60 deg due to the expanding flow). Figure \ref{fs-LCD} (right) shows the re-orientation of the magnetic field lines due to the HBI within $\sim$100 kpc at 1.27 Gyr, which is in qualitative agreement with the 3-D simulation results of \citet{2009ApJ...703...96P} and \citet{2010ApJ...713.1332R}. Also noted is the qualitative similarity of our solution for the heat flux factor in Fig. \ref{fs-LCD} (left) with that obtained by \citet{2009ApJ...703...96P} and \citet{2010ApJ...713.1332R} in three dimensions before saturation (to the value of 0.07). At about 1.3 Gyr the velocity at $R=$20 kpc begins to increase as the first expansion wave is reflected off the outer boundaries of our physical domain and reaches the near-core region. Beyond this time the magnetic field orientation with respect to the radial position vector, the heat flux factor, and the temperature are driven by the hydrodynamics of the flow. 

\begin{figure*}%\includegraphics[trim = <trim from left> <from bottom> <from right> <from top>, clip, <other options>]{<image filename>}
\includegraphics[trim = 13mm 1mm 5mm 5mm, clip, scale=0.33]{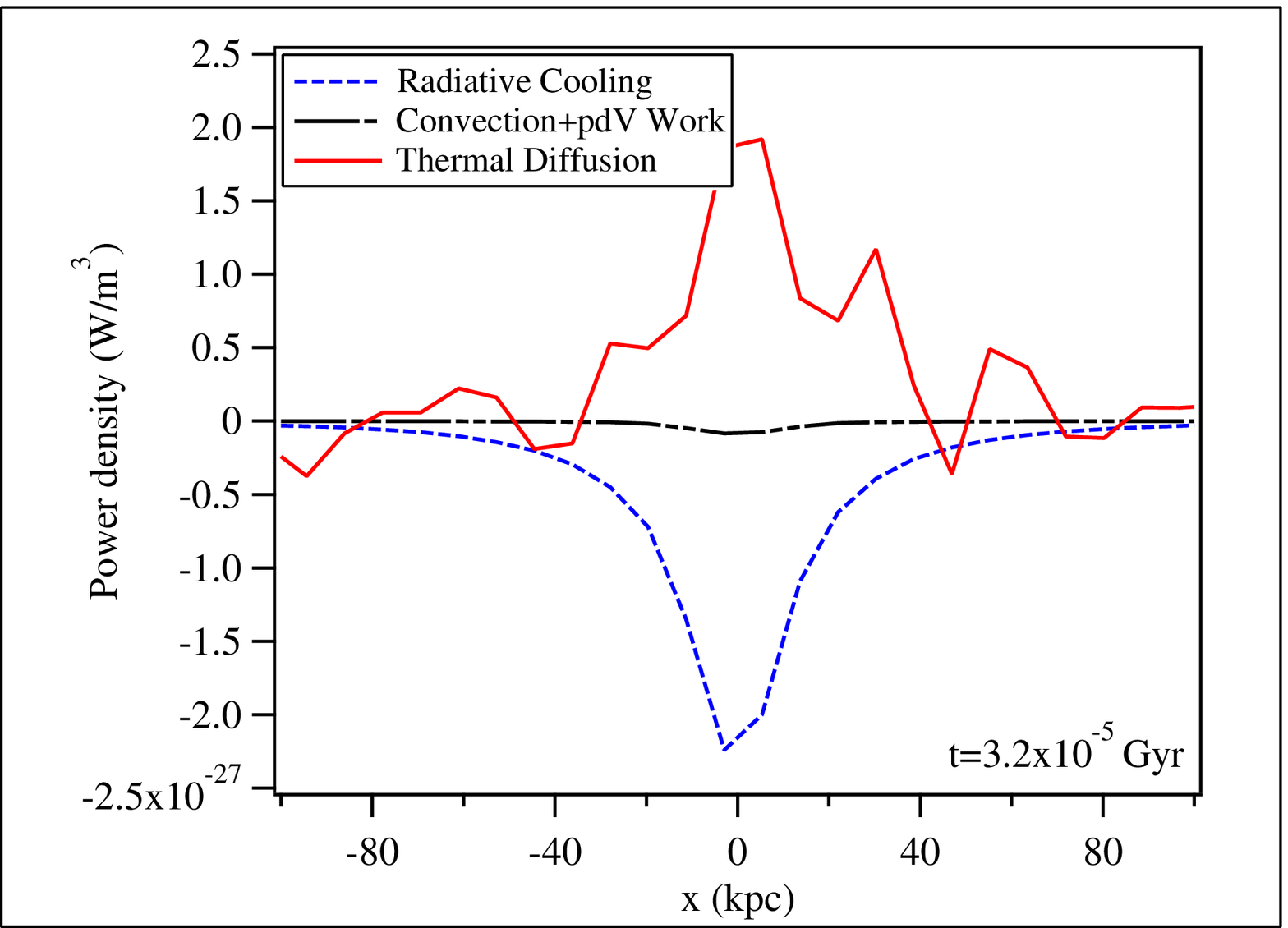}
\includegraphics[trim = 13mm 1mm 5mm 5mm, clip, scale=0.33]{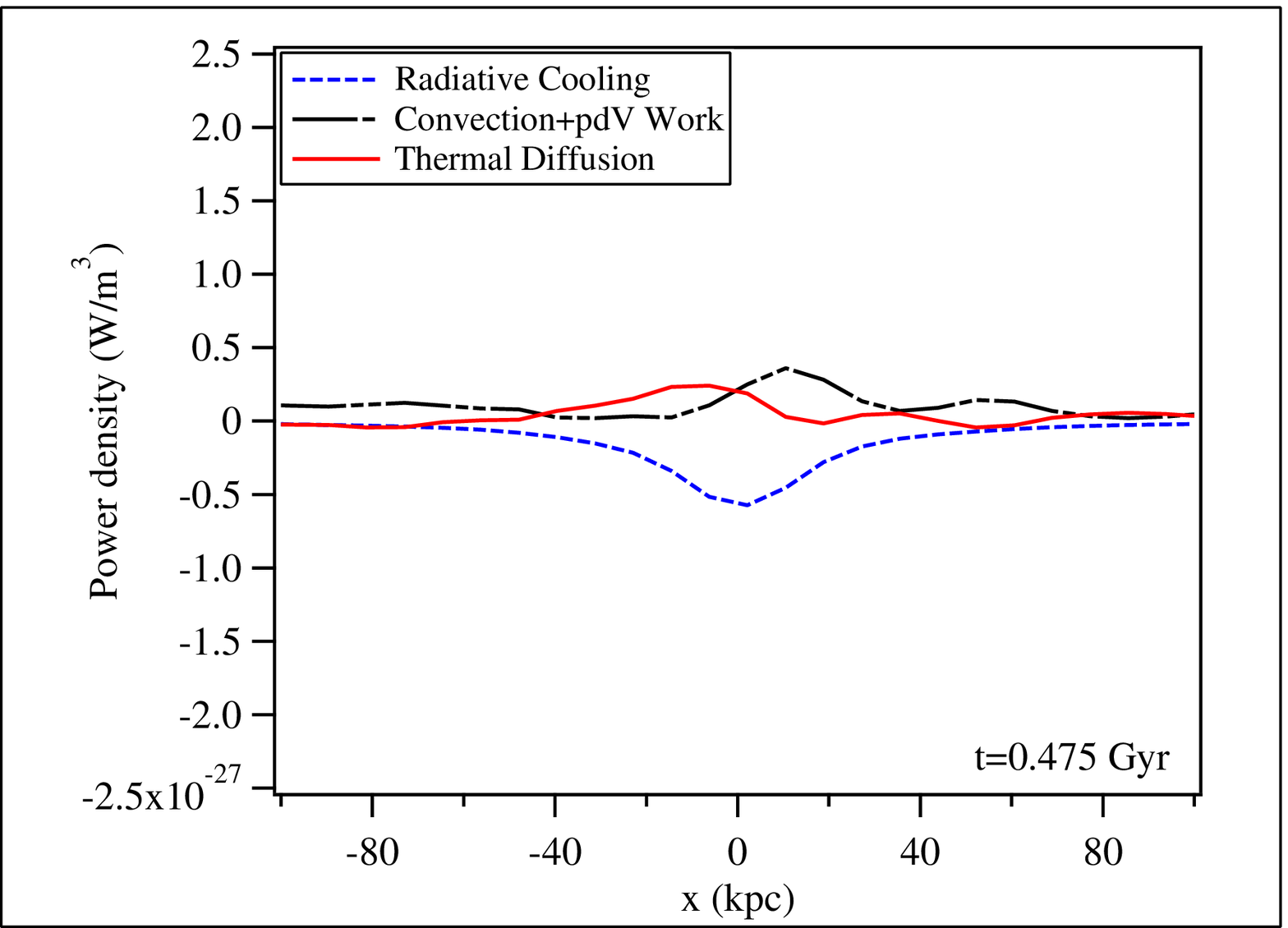}
\includegraphics[trim = 13mm 1mm 5mm 5mm, clip, scale=0.33]{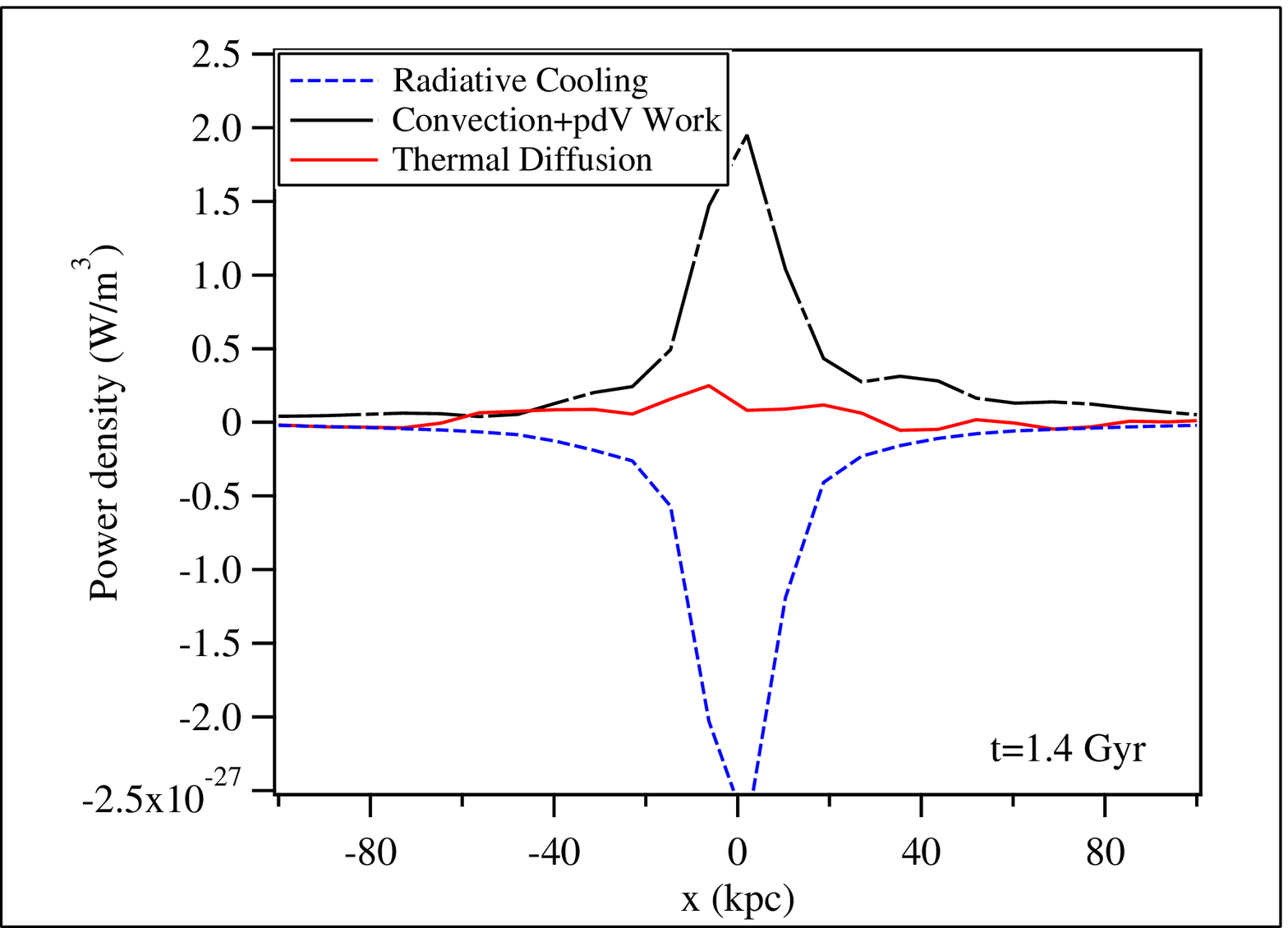}
\caption{Comparison of the contributions to the ICM energy balance from the convection and pdV work, radiation, and anisotropic thermal conduction. The comparisons are for the case of the LCD initial profile in planar geometry, employing CRM simulation physics and a 1 $\mu$G turbulent magnetic field. The slice plots are along x at y=0. The computational mesh consisted of $128^2$ square elements yielding a minimum size of 7.8 kpc for each element.}
\label{PowerDenSlic}
\end{figure*}

A closer look at the various terms in the energy equation shows the diminishing importance of thermal conduction near the core in the early energy balance of the system, with the convective and radiative contributions almost fully driving the system by $\sim$1.4 Gyrs. Figure \ref{PowerDenSlic} compares the power densities for radiative cooling $\Phi_{eR}$, thermal conduction $\nabla\cdot\textbf{\textit{q}}$ and the convective terms -$(\rho \textbf{\textit{v}}\cdot\nabla \epsilon+p\nabla \cdot \textbf{\textit{v}})$, at three different times along the $y=$0 axis. In all cases we found that $p\nabla \cdot \textbf{\textit{v}}$ is the dominant term over $\rho \textbf{\textit{v}}\cdot\nabla \epsilon$. Also, because the initial conditions imposed zero velocity, the plot at $t=3.2\times 10^{-5}$ Gyr simply reflects that the evolution of the internal energy of the system is initially driven by thermal conduction and radiation, but by $\sim$1.4 Gyrs temperature gradients near the core are found to be almost completely smoothed out. 
%%%%%%%%%%%%%%%%%%%%%%%%%%%%%%%%%%%%%%%%%%%
%%%%%%%%%%%%%%%%%%%%%%%%%%%%%%%%%%%%%%%%%%%
\begin{figure*}
\includegraphics[trim = 1.5mm 1mm 5mm 2mm, clip, scale=0.44]{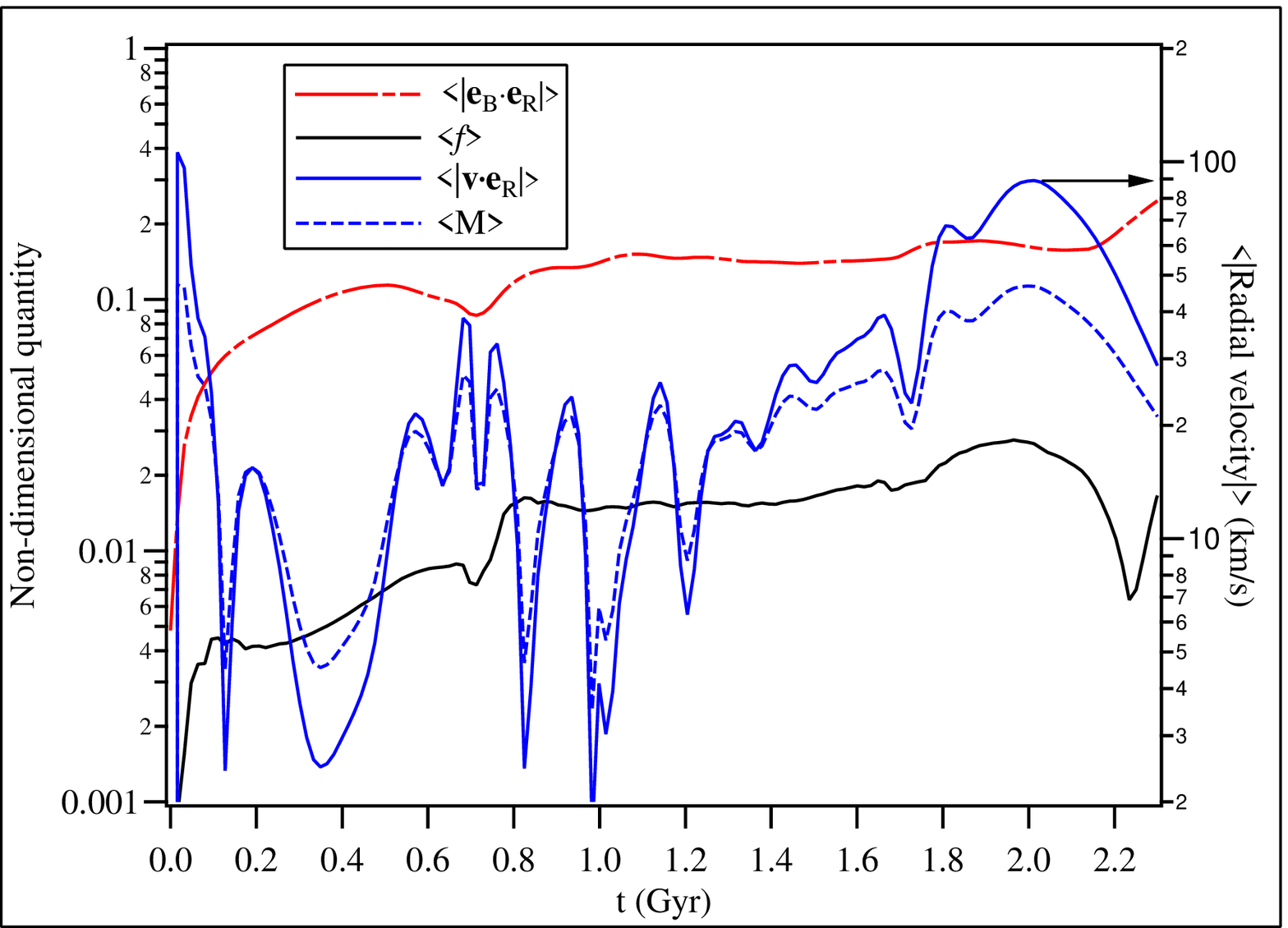}
\includegraphics[trim = 1.5mm 1mm 5mm 2mm, clip, scale=0.44]{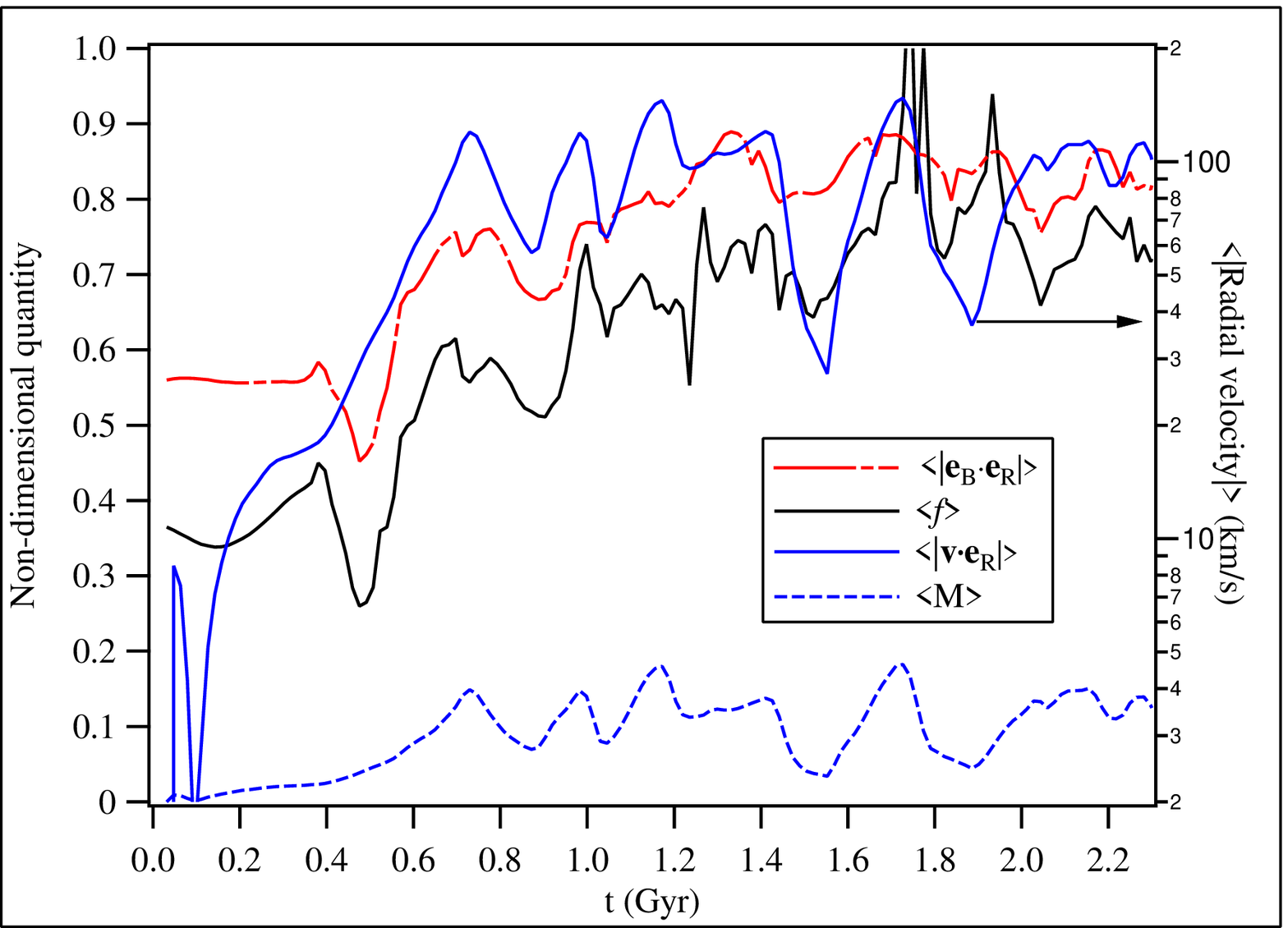}
 \caption{Evolution of $\langle\left|\bmath{\hat{e}}_B \cdot \bmath{\hat{e}}_R\right|\rangle$, $\langle\textit{f}\rangle$, $\langle\left|\textbf{\textit{v}} \cdot \bmath{\hat{e}}_R\right|\rangle$ and $\langle\left|\textbf{\textit{v}} \cdot \bmath{\hat{e}}_R\right|\rangle/\langle c_s\rangle$, averaged over the area of a disc of radius $R$=20 kpc ($\pm \Delta$). The cases employed CRM simulation physics in planar geometry. Left: Results starting with the LCD initial condition for the density and temperature and a rotational initial magnetic field with maximum magnitude of 1 $\mu$G.  The velocity flow field here fluctuates between radially inward and outward directions after the first expansion at t$\sim$0.1 Gyr (note that we plot the absolute value of the radial velocity). Right: Results starting with the HCD initial conditions for the density and temperature and a turbulent initial magnetic field with maximum magnitude of 1 $\mu$G. The flow field is directed inward after the first expansion at t$\sim$0.1 Gyr (see also Fig. \ref{2DHCD-trb}).}
\label{fs-azm-LCD}
\end{figure*}
%%%%%%%%%%%%%%%%%%%%%%%%%%%%%%%%%%%%%%%%%%%
\begin{figure*}
\includegraphics[trim = 1mm 1mm 1mm 1mm, clip, scale=0.37]{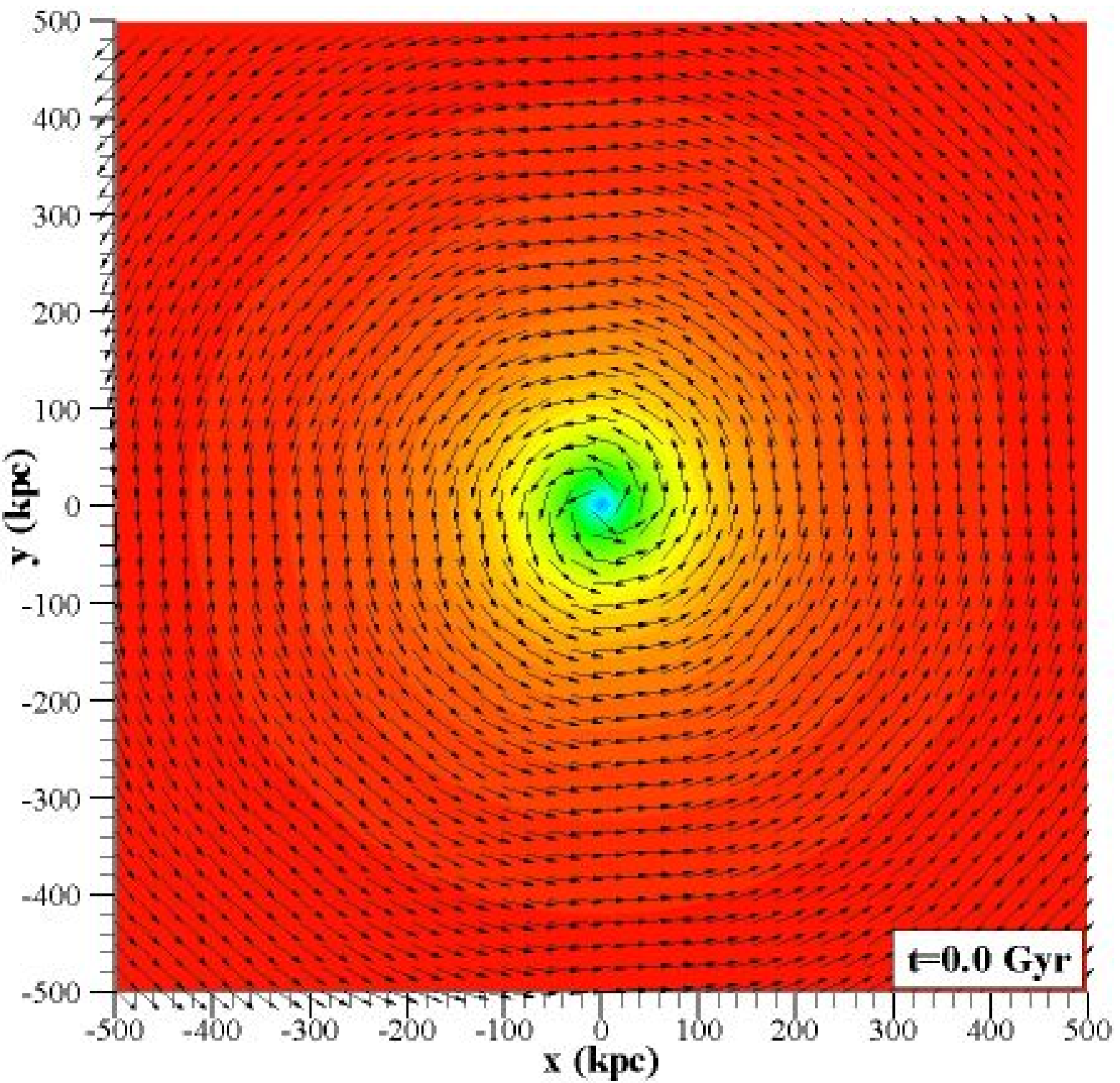}
\includegraphics[trim = 1mm 1mm 1mm 1mm, clip, scale=0.495]{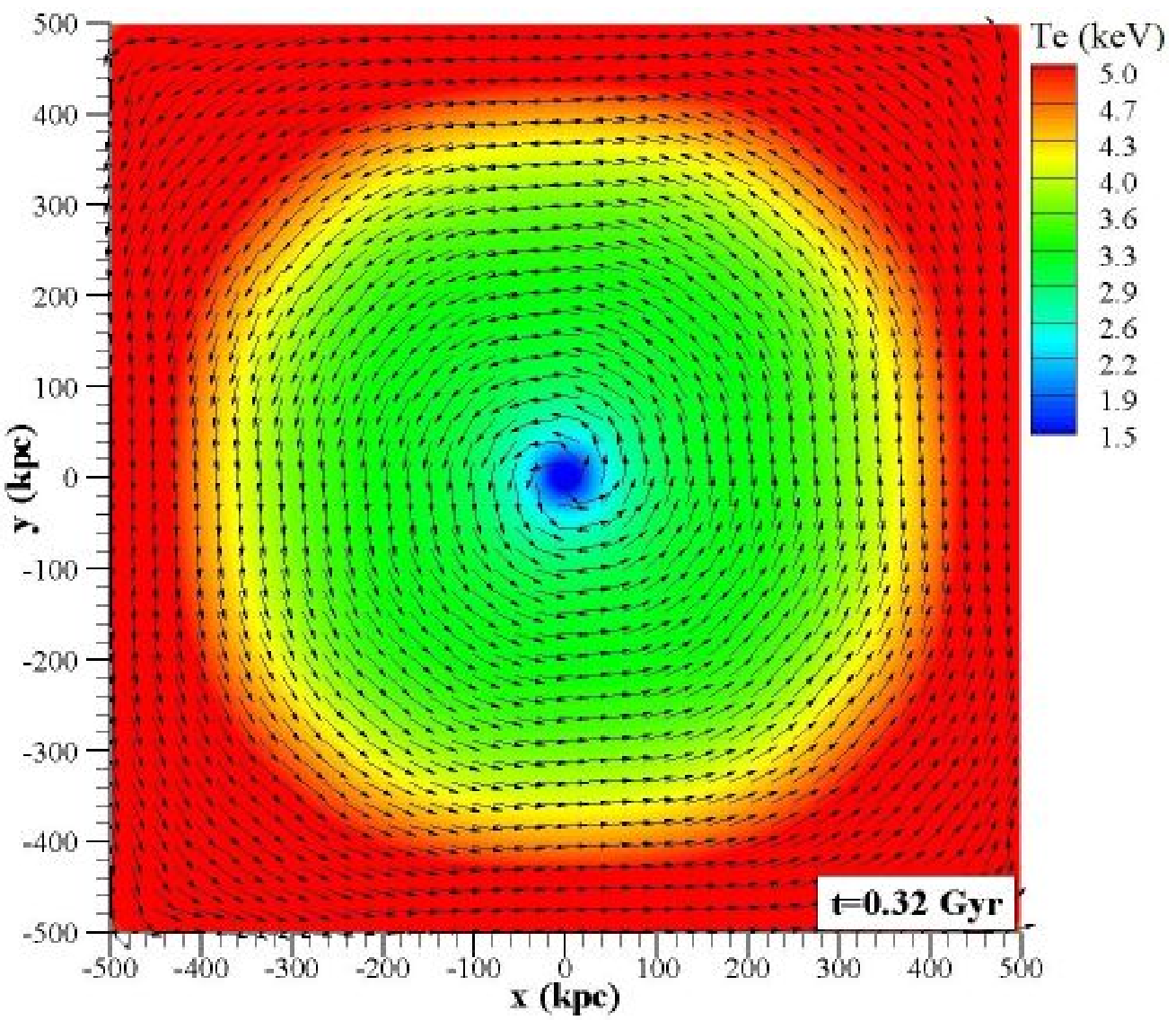}
\caption{Evolution of the temperature (in eV) and magnetic (unit vector) field for the case of the LCD initial profile and a rotational initial magnetic field with peak magnitude of a 1 $\mu$G. CRM simulation physics were used in planar geometry.}
\label{2DLCD-azm}
\end{figure*}

The diminishing effect of anisotropic thermal conduction is even more evident in the case of a rotational initial magnetic field, which is found to reduce on-average the heat flux to the core within the time interval of interest by more than one order of magnitude compared to the case of a turbulent initial field (see Fig. \ref{fs-LCD} (left)). The evolution of the pertinent average parameters is shown Fig. \ref{fs-azm-LCD} (left). The rotational magnetic field remains approximately orthogonal to the radial unit vector; the angle between them does not fall below $\sim$78 deg. Yet, this has little effect on the evolution of the temperature during this time. This may seem counterintuitive at first since the characteristic times associated with the smoothing of temperature gradients in the ICM by thermal conduction are larger in the rotational-field case as the effective heat flux factor is significantly lower compared to the turbulent field (equation \ref{eqn:HBIGrowth} gives $\tau_q \sim$12 Gyr for \textit{f}=0.02). However, thermal conduction does not drive the temperature in these cases. The characteristic time for temperature changes induced by the pressure work on the fluid may be estimated by
\begin{equation}
\frac{3}{2}n_e k_B \frac{\partial T_e} {\partial t}=-p\nabla \cdot \textbf{\textit{v}}
\end{equation}
yielding
\begin{equation}
\tau_p \approx \frac{3}{2}\frac{\Delta T_e}{T_e}\frac{\ell}{\Delta \textit{v}} .
\end{equation}
For the induced conditions in our numerical experiments with the LCD profiles $\tau_p \approx$0.13 Gyr ($<<\tau_q$). Figure \ref{2DLCD-azm} depicts the evolution of the temperature showing the smoothing of the temperature gradients despite the preservation of a highly-rotational magnetic field within the time interval of interest. Thus, under the induced conditions $p\nabla \cdot \textbf{\textit{v}}$ dominates over thermal conduction.

\begin{figure}
\includegraphics[trim = 1mm 1mm 1mm 1mm, clip, scale=0.65]{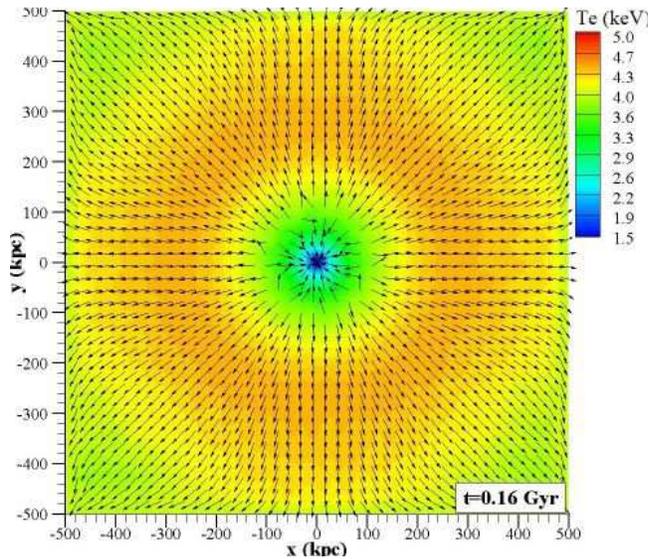}
\caption{Temperature (in eV) and velocity (unit vector) field for the case of the HCD initial profile and a turbulent initial magnetic field with peak magnitude of a 1 $\mu$G and CRM simulation physics in planar geometry.}
\label{2DHCD-trb}
\end{figure}

\begin{figure}
\includegraphics[trim = 1.5mm 1mm 5mm 2mm, clip, scale=0.47]{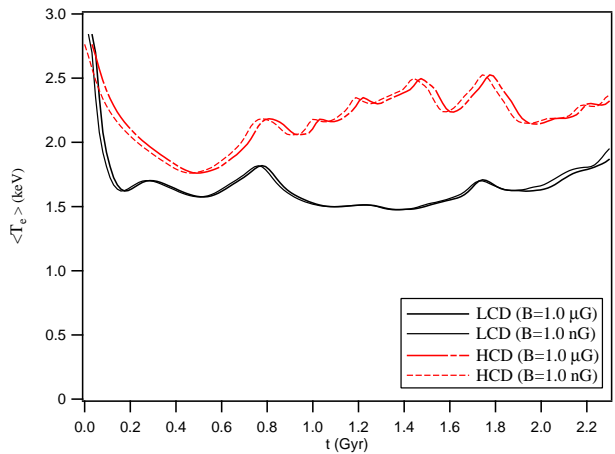}
 \caption{Comparisons between the average temperature for the LCD and HCD profiles starting with a turbulent initial magnetic field with maximum magnitude of 1 nG and 1 $\mu$G. All cases employed CRM simulation physics in planar geometry.}
\label{CompareZNwithoPoly2}
\end{figure}

The 2-D axisymmetric simulations in Section \ref{subsec:AxisymSims} with only radiation and thermal conduction demonstrate a strong sensitivity of the solution on the assumed initial core density, with the HCD profile yielding a collapse of the core at values of $\textit{f}$ for which the LCD profile yielded no cooling. The results of the cosine of the magnetic field angle, heat flux factor, radial velocity and Mach number are depicted in Fig. \ref{fs-azm-LCD} (right) for the HCD profiles. The hydrodynamics of the flow here are significantly different compared to the LCD case. The induced hydrostatic non-equilibrium state for the ICM encompasses higher energy density in the system but at a reduced pressure gradient. The resulting expansion is weaker compared to the LCD case (10 km/s wave front speed in HCD versus 100 km/s in LCD) and the gravitational force is able to turn the flow at $R=$20 kpc inward relatively fast, within about 0.1-0.2 Gyr, and the plasma beyond this time continues to flow radially inward. This is in contrast to the LCD case with a rotational magnetic field where the velocity field fluctuates between inward and outward directions (note that we plot the absolute value of the radial velocity). The temperature contours at t=0.16 Gyr are shown in Fig. \ref{2DHCD-trb}. Thus, the response of the temperature in Fig. \ref{CompareZNwithoPoly2} and of the non-dimensional quantities in Fig. \ref{fs-azm-LCD} (right) responded considerably faster than in the LCD case. The solutions with a peak strength for the initial magnetic field of 1 nG show only minor differences with the 1-$\mu$G cases.
%%%%%%%%%%%%%%%%%%%%%%%%%%%%%%%%%%%%%%%%%%%
%%%%%%%%%%%%%%%%%%%%%%%%%%%%%%%%%%%%%%%%%%%
%%%%%%%%%%%%%%%%%%%%%%%%%%%%%%%%%%%%%%%%%%%

\section{Summary and Discussion}\label{sec:Discussion}
The works of \citet{2003ApJ...582..162Z}, \citet{2008ApJ...681..151C} and \citet{2009ApJ...703...96P} have been particularly influential in formulating the different cases presented in this paper. Our 2-D axisymmetric simulations reproduce closely the effective heat flux factor of \citet{2003ApJ...582..162Z} under similar assumptions for A2199, when thermal conduction and (bremsstrahlung) radiation are the only physics allowed in the energy balance of the cluster, and when the electron number density and temperature profiles closely resemble the authors' ``best-fit'' solution; we have termed this solution as the ``LCD profile.'' 

Because the LCD profile underestimates the \textit{Chandra} density observations near the core by approximately 35\% (albeit in excellent agreement with the data beyond $\sim$10 kpc), we performed a second set of simulations with an initial profile that was in agreement with the observed electron density near the cluster center but overestimated the data at larger distances. Differences in the temperature profiles between the two cases were less than 5\% for R$<$200 kpc. Because the ratio of characteristic time for thermal conduction over the time for radiative cooling is strongly dependent on the electron density, $\tau_q/\tau_{\Phi} \propto {n_e}^2$ (assuming everyting else fixed), the goal of these simulations was to determine what effect such an increase would have on the thermal balance in the cluster. By comparion to the LCD case we found that the ICM collapsed catastrophically in the HCD case for the same values of the heat flux factor \textit{f}. Our conclusions from this series of idealized simulations regarding the importance of the core density in the evolution of the ICM are in agreement with the conclusions of \citet{2008ApJ...681..151C}, who illustrated the dependence of the ICM collapse rate on the value of the plasma density at the center. A turbulent initial magnetic field that was held fixed throughout the simulation also led to the collapse of the ICM for both initial density profiles (LCD and HCD). In this series of simulations the anisotropy of the conduction heat flux was accounted for by invoking the approximations of \citet{1965RPP} for the thermal conductivity coefficients. The magnetic field was generated by a method proposed by \citet{2010ApJ...713.1332R}, and was held frozen to emulate an extreme case in which galaxy motion maintains the turbulent field in the absence of any flow dynamics in the ICM.

Since it remains largely infeasible to perform fully-consistent cosmological simulations of clusters that account for all pertinent physics and scale lengths, most often in MHD simulations the thermodynamic state and magnetic field topology are prescribed as the ``initial conditions'' of the ICM cluster and the simulations then proceed to seek of mechanisms that maintain thermal balance in the cluster. Such simulations do not always begin with an ICM in thermal balance but all studies we have reviewed imposed hydrostatic equilibrium at t=0. Referring to Fig. \ref{InitialProfiles}, it is interesting to note that the trend of increasing density near the core ($R\leq$10 kpc) based on the observed data appears to be one that crosses the two hydrostatic-equilibrium solutions, namely the LCD and HCD profiles. This suggests that, at least near the core of A2199, deviations from such equilibria may be at play. In fact, \citet{2006MNRAS.371L..65S} hypothesize the presence of a relatively weak ($M\sim$1.5) isothermal shock in this region. The MHD simulations in 2-D planar geometry we performed here were intended to serve as a preliminary series of numerical experiments to assess possible deviations from hydrostatic equilibrium in connection with transient AGN activity and/or remnant dynamical activity from the early formation of the galaxy cluster. To span a wide range of magnetic field effects we employed initial strengths and topologies similar to those imposed by \citet{2009ApJ...703...96P} and \citet{2009ApJ...704..211B}. The 2-D results now also form the basis for near-future 3-D simulations with MACH3. 

We were particularly interested in the initial response of the plasma near the core ($t<$1.3 Gyr, $R=$20 kpc) because radiative cooling and the HBI have been shown to lead to a runaway catastrophe in A2199 in this region in less than 3 Gyr (\citealt{2009ApJ...703...96P, 2010ApJ...713.1332R}). Moreover, it has been proposed that mass and momentum transport from low-speed winds ($\sim$10,000 km/s) associated with AGN outbursts, can have a dramatic effect on the growth of the central supermassive black hole (\citealt{2010arXiv1004.2923O}) and, in turn, on the AGN-ICM coupling. We deliberately imposed in the initial thermodynamic state of the ICM a deviation from hydrostatic equilibrium, thereby producing subsonic expansion flow upon initiation of the simulation. By inducing a non-zero velocity field early in the simulation we allowed for (thermal) pressure work to contribute to the energy balance of the system. In fact, under the induced conditions, pressure work eventually overwhelmed the deleterious effects of the magnetic field, which in the idealized hydrostatic case would have shut off the main supplier of heat to the region (in the absence of an AGN sources), namely thermal conduction. It is noted that only the scalar (thermal) pressure was included in this work; its importance in the presence of a subsonic velocity field and a magnetic field in this region suggests that the full pressure tensor may have to be accounted for as proposed by \citet{2010arXiv1003.2719K}). In all three cases studied with different initial magnetic field, and for both initial density profiles (HCD and LCD), we obtained similar results by 2-D MHD simulations about the sustainment of the temperature near the core.

\section{Conclusions}\label{sec:Conclusions}
In this paper we presented results from a series of numerical experiments in two dimensions for the A2199 galaxy cluster. The simulations employed a relatively wide combination of initial conditions and simulation physics. It is acknowledged that the 2-D physical domains did not permit us to address a wide range of possible physics in clusters that are inherently 3-D, and the propositions presented herein shall be re-evaluated in three dimensions. Nevertheless, the number of cases performed in this study yielded constructive insight that now also forms the basis of our near-future simulations with MACH3. 

We chose A2199 largely because observed data for this cluster exist at a relatively small radius from its center ($R\sim$1 kpc). As a consequence, our idealized 2-D axisymmetric simulations aimed to quantify the sensitivity of the ICM thermal balance on the plasma density profiles with less ambiguity compared to other clusters for which such data does not exist. The computed differences in the heat flux factor for the two profiles we used were not insignificant. Therefore, it appears it would be instructive that MHD parametric studies, performed to asses the sensitivity of proposed dominant physics in the ICM and its sustainment/collapse (e.g. HBI, MTI, AGN heating, turbulence stirring), extended their range to include higher values of the plasma density near the cluster center. 

The results of the 2-D planar simulations demonstrate that imbalances in the system's hydrostatic equilibrium, inducing relatively weak flow dynamics, may overcome the effects of the magnetic field on the diffusion of heat from the outer regions of the cluster. This marginalizes the significance of anisotropic thermal conduction over transient flow events in the ICM. Such events may possibly be associated with AGN pulsations at the cluster center, with characteristic dynamical times that are hundreds of times shorter than the Hubble time.

\section*{Acknowledgments}
The authors wish to thank Jeremiah Ostriker and Matt Kunz for their useful comments. We are also grateful to Moustafa T. Chahine for his support through the Innovative Spontaneous Concepts Research and Technology Program. This work was carried out at the 
Jet Propulsion Laboratory, California Institute of Technology, under a contract with the National Aeronautics 
and Space Administration, and made extensive use of the NASA Astrophysics Data System and {\tt arXiv.org} preprint server.\\
\copyright 2010. All rights reserved.
%
%
%%%%%%%%%%%%%%%%%%%%%%%%%%%%%%%
%
%

%
%
%
%
%
\appendix
\section[]{The full MHD equations of MACH}\label{sec:Appendix}
\subsection{Conservation equations}\label{subsec:AppxEquations}
The MACH codes solve equations (\ref{MACHcontinuity})-(\ref{MACHinduction}), the time-dependent, compressible, single-fluid, multi-temperature resistive MHD conservation laws, on a mesh that is composed of arbitrary hexahedral cells. 
\begin{eqnarray}
%Fluid continuity
\frac{\partial \rho}{\partial t}&=&-\nabla\cdot\left(\rho \textbf{\textit{v}}\right)
\end{eqnarray}
\label{MACHcontinuity}
\begin{eqnarray}
%Fluid momentum
\rho\frac{\partial \textbf{\textit{v}}}{\partial t}&=&-\rho \textbf{\textit{v}}\cdot\nabla \textbf{\textit{v}}-\nabla\left(p+Q+\epsilon_R/3\right)\nonumber \\
&+&\textbf{\textit{J}}\times\textbf{\textit{B}}+\rho \textbf{\textit{g}}
\end{eqnarray}
\label{MACHmomentum}
\begin{eqnarray}
%Electron specific internal energy
\rho\frac{\partial \epsilon_e}{\partial t}&=&-\rho \textbf{\textit{v}}\cdot\nabla \epsilon_e-p_e\nabla \cdot \textbf{\textit{v}}+\nabla\cdot\left(\bmath{\bar{\bar{\kappa}}}_e\cdot \nabla T_e\right)-\Phi_{eR}\nonumber \\
&+&\textbf{\textit{J}}\cdot\left(\bmath{\bar{\bar{\eta}}}\cdot \textbf{\textit{J}}\right)-\textbf{\textit{J}}\cdot\frac{\nabla p_e}{en_e}-\rho C_{V}\left(T_e-T_i\right)/\tau_{ei}
\label{MACHeenergy}
\end{eqnarray}
\begin{eqnarray}
%Ion specific internal energy
\rho\frac{\partial \epsilon_i}{\partial t}&=&-\rho \textbf{\textit{v}}\cdot\nabla \epsilon_i-\left(p_i+Q\right)\nabla\cdot \textbf{\textit{v}}+\nabla\cdot\left(\bmath{\bar{\bar{\kappa}}}_i\cdot \nabla T_i\right)\nonumber\\
&+&\rho C_{V}\left(T_e-T_i\right)/\tau_{e}
\label{MACHienergy}
\end{eqnarray}
\begin{eqnarray}
%Radiation energy density
\frac{\partial \epsilon_R}{\partial t}&=&-\rho \textbf{\textit{v}}\cdot\nabla \epsilon_R-{4/3}\epsilon_R\nabla\cdot \textbf{\textit{v}}\nonumber \\
&+&\nabla\cdot\left(\rho \chi_r\nabla \epsilon_R\right)+\Phi_{eR}
\label{MACHrenergy}
\end{eqnarray}
\begin{eqnarray}
%Magnetic induction
\frac{\partial \textbf{\textit{B}}}{\partial t}&=&\nabla\times\left(\textbf{\textit{v}}\times \textbf{\textit{B}}\right)-\nabla\times\left(\bmath{\bar{\bar{\eta}}}\cdot \textbf{\textit{J}}\right) \nonumber\\
&-&\nabla\times\left(\frac{\textbf{\textit{J}}\times\textbf{\textit{B}}}{en_e}\right)+\nabla\times\left(\frac{\nabla p_e}{en_e}\right)
\label{MACHinduction}
\end{eqnarray}
In equations (\ref{MACHcontinuity})-(\ref{MACHinduction})
 $\rho$ is the mass density of the fluid,
 $\textbf{\textit{v}}$ is the fluid velocity,
 $\textbf{\textit{B}}$ is the magnetic induction,
 $\epsilon_{e/i}$ is the electron/ion specific internal energy,
 $\epsilon_R$ is the radiation energy density, and is related to the radiation temperature $T_R$ via Stefan's constant \textit{a}: $\epsilon_R=a T_R^4$. Stefan's constant is related to the Stefan-Boltzmann constant $\sigma =a c/4$ where $c$ is the speed of light in vacuum.
 $T_{e/i}$ is the electron/ion temperature,
 $p_{e/i}$ is the electron/ion thermal pressure,
 $Q$ is the artificial numerical compressional viscosity,
 $\textbf{\textit{J}}$ is the current density,
 $n_e$ is the electron number density,
 $C_{V}$ is the specific heat,
 $\tau_{ei}$ is the electron-ion equilibration time, 
 $e$ is the magnitude of the charge of an electron and $\mu_0$ is the permeability of free space.
 The electrical resistivity tensor is $\bmath{\bar{\bar{\eta}}}$ and 
 $\bmath{\bar{\bar{\kappa}}}_{e/i}$ is the electron/ion thermal conductivity tensor.

Energy is conserved separately for electrons and ions. Models for radiation emission (thin limit), equilibrium conduction (thick limit) as well as non-equilibrium (flux-limited) radiation conduction also exist. The \textit{sesame} tables serve as tabular equations of state for more than one hundred materials. The library is created and maintained by two groups at the Los Alamos National Laboratory (EOS and conductivity tables are the work of T-1; opacity tables are created and maintained by T-4). Although tabular values for the electrical and thermal conductivities are often available, the default models of our MACH version use the Spitzer-Braginskii conductivities parallel and perpendicular to the magnetic field separately for electrons and ions. Most of the physical processes are time-advanced in an implicit fashion that allows the Courant condition to be exceeded without exciting numerical instabilities. The spatial derivatives are obtained by integrating over finite volumes. The clean-up procedure applied to $\textbf{\textit{B}}$ to maintain its divergence at an acceptably small level (solenoidal condition) uses a successive over-relaxation solver to iterate on the in-plane components toward $\nabla \cdot \textbf{\textit{B}}=0$ by solving for the scalar potential $\phi$ (elliptic correction since a Poisson's equation is solved), and is based on the method proposed by \citet{1980JCoPh..35..426B}. 

In the last three decades MACH2 has been employed to study a variety of plasma problems some of which include plasma opening switches (\citealt{1987ITPS...15..766B}; \citealt{1987ITPS...15..760D}), gas and solid density z-pinch implosion physics (\citealt{1995PhRvL..74...98D}; \citealt{1995ITPS...23...83S}), very high-power plasma source designs (\citealt{2002JPP..18..146M}),  magnetic nozzles (\citealt{2002JPP..18..152M}), and a variety of plasma thrusters (\citealt{2000JPP..16..887M}; \citealt{2000mfss.conf..353T}; \citealt{2007JAP...102j3301M}).

\begin{figure}
\includegraphics[trim = 1.0mm 0mm 4mm 2mm, clip, scale=0.55]{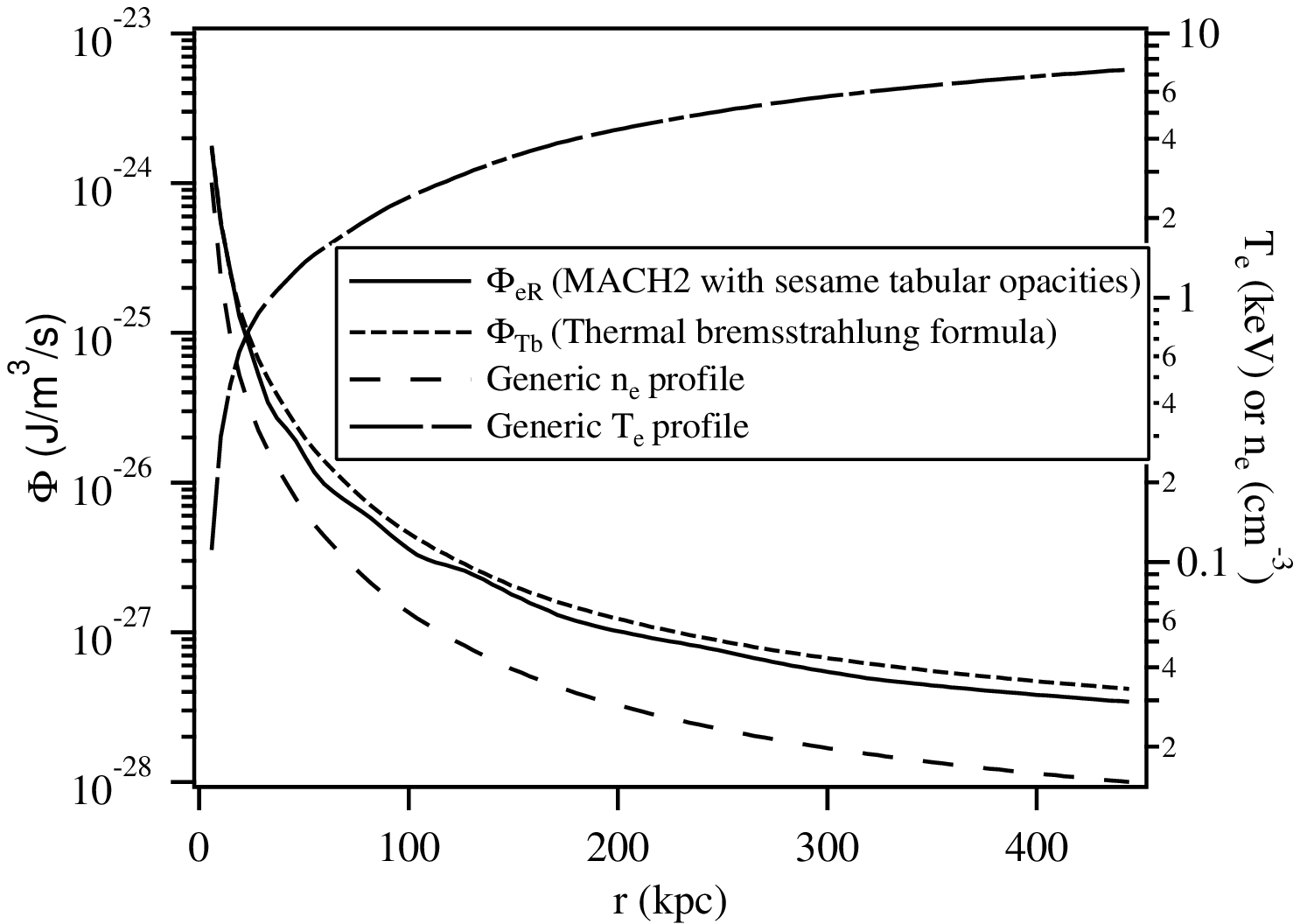}
 \caption{Comparison of the radiation emission rates based on the MACH2 tabular opacities and the formula by \citet{1979rpa..book.....R} with a Gaunt factor of 1.5.}
 \label{BremRadiationProfile}
\end{figure}

MACH is a three-temperature code, that is, the temperature of the radiation field $T_R$ may be different from $T_{e/i}$. In this case the radiation conduction equation (\ref{MACHrenergy}) is solved where both Planck (optically-thin limit) $\chi_p$ and Rosseland (optically-thick limit) $\chi_r$ mean opacities must be supplied. The radiation coupling term is in general $\Phi_{eR}=a c \rho\chi_p(T_e^4-T_R^4)$. Figure \ref{BremRadiationProfile} compares the radiation emission rates based on the MACH2 tabular opacities and the formula by \citet{1979rpa..book.....R} with a Gaunt factor of 1.5 (note the factor ranges 1.1-1.5).
\subsection{Anisotropic Thermal Conduction Flux in MACH2} \label{subsec:AppxConductionFlux}
In general, the conductive heat flux for electrons in a magnetic field (in the absence the thermoelectric effect) consists of parallel, perpendicular and transverse components as follows:
\begin{eqnarray}
\bmath{q}_{e}=\bmath{q}_{e//}+\bmath{q}_{e \bot}+\bmath{q}_{e \wedge}
\end{eqnarray}
where
\begin{eqnarray}
\bmath{q}_{e//}&=&-\kappa_{e//} \nabla_{//}T_e \nonumber\\
\bmath{q}_{e\bot}&=&-\kappa_{e\bot} \nabla_{\bot}T_e \nonumber\\
\bmath{q}_{e\wedge}&=&-\kappa_{e\wedge} \bmath{\hat{e}}_B \times \nabla T_e
\end{eqnarray}
and $\nabla_{//}\equiv \bmath{\hat{e}}_B \bmath{\hat{e}}_B \cdot \nabla$, $\nabla_{\bot}\equiv \left({\bar{\bar{\textbf{I}}}}-\bmath{\hat{e}}_B\bmath{\hat{e}}_B\right) \cdot \nabla$. The conductivity coefficients are given by
\begin{eqnarray}
\kappa_{e//}&=&3.16 \kappa_{e0} \nonumber\\
\kappa_{e\bot}&=&\frac{4.66 \Omega_e^2+11.92}{\Omega_e^4+14.79 \Omega_e^2+3.77}\kappa_{e0}\nonumber\\
\kappa_{e\wedge}&=&\frac{\Omega_e \left(2.5 \Omega_e^2+21.67\right)}{\Omega_e^4+14.79 \Omega_e^2+3.77}\kappa_{e0},
\label{BraginskiiCoeffs}
\end{eqnarray}
and are the approximations of \citet{1965RPP} for the case of a fully-ionized plasma, with $\kappa_{e0}=k_B p_e \tau_e/m_e$. Thermal diffusion is solved for implicitly in MACH with an iterative Jacobi method. The rate of convergence of the Jacobi iteration may be increased by multigrid methods (e.g. \citealt{1973LNP....18...82B}), which we have employed in the present MACH2 simulations.

\label{lastpage}

\end{document}